\documentclass[twocolumn,showpacs,preprintnumbers,amsmath,amssymb,aps]{revtex4}
\usepackage{epsfig}
\pdfoutput=1

\usepackage[T1]{fontenc}
\usepackage[latin9]{inputenc}
\usepackage{amsmath,amssymb,amsthm,amsfonts}
\usepackage{slashed}
\usepackage{graphicx}
\usepackage{color,rotating}
\usepackage{dsfont}
\usepackage{setspace}
\usepackage{verbatim}
\usepackage{fancyhdr}
\newcommand{\sumint}{\int\hspace{-4.8mm}\sum}
\newcommand{\step}{\vspace{.5em}}
\newcommand{\smallstep}{\vspace{.06em}}

\newcommand*{\pd}{\partial}

\newcommand*{\mn}{{\mu\nu}}

\def\di{\displaystyle}

\def\bg{\begin{eqnarray}\begin{array}{rcl}\displaystyle}
\def\eg{\end{array} &\di    &\di   \end{eqnarray}}
\def\bm#1{\begin{eqnarray}\begin{array}{#1}\di}
\def\bmo#1{\begin{eqnarray*}\begin{array}{#1}\di}
\def\bml#1#2{\begin{eqnarray}\begin{array}{#1}\label{#2}\di}
\def\bgo{\begin{eqnarray*}\begin{array}{rcl}\displaystyle}
\def\ego{\end{array} &\di    &\di \nonumber  \end{eqnarray*}}
\def\btensor#1#2{\renew\left#1\begin{array}{#2}\di}
\def\brtensor#1#2#3{\ren#3\left#1\begin{array}{#2}}
\def\botensor#1#2{\renew\left#1\begin{array}{#2}}
\def\etensor#1{\end{array}\right#1}

\def\eq#1{(\ref{#1})}
\def\Eq#1{Eq.~(\ref{#1})}

\def\Tr{{\rm Tr}}

\def\s0#1#2{\mbox{\small{$ \frac{#1}{#2} $}}}
\def\0#1#2{\frac{#1}{#2}}

\def\CO{{\mathcal O}}

\def\CT{{\mathcal T}}

\def\ra{\rightarrow}

\def\ren#1{\renewcommand{\arraystretch}{#1}}

\def\renew{\renewcommand{\arraystretch}{1}}

\newcommand{\beq}{\begin{equation}}
\newcommand{\eeq}{\end{equation}}
\newcommand{\beqa}{\begin{eqnarray}}
\newcommand{\eeqa}{\end{eqnarray}}
\newcommand{\nn}{\nonumber}
\newcommand{\tn}{\textnormal}

\begin{document}

\title{Yang-Mills correlation functions at finite temperature}
 
\vspace{1.5 true cm}
 
\author{Leonard Fister}

\affiliation{Institut f\"ur Theoretische Physik, 
Universit\"at Heidelberg, 
Philosophenweg 16, 69120 Heidelberg, Germany}

\affiliation{ExtreMe Matter Institute EMMI, GSI Helmholtzzentrum f\"ur Schwerionenforschung mbH, Planckstr. 1, D-64291 Darmstadt, Germany}

\author{Jan M. Pawlowski}

\affiliation{Institut f\"ur Theoretische Physik, 
Universit\"at Heidelberg, 
Philosophenweg 16, 69120 Heidelberg, Germany}

\affiliation{ExtreMe Matter Institute EMMI, GSI Helmholtzzentrum f\"ur Schwerionenforschung mbH, Planckstr. 1, D-64291 Darmstadt, Germany}

\begin{abstract}
  {We put forward a continuum approach for computing finite
    temperature correlation functions in Yang-Mills theory. This is
    done in a functional renormalisation group setting which allows
    for the construction of purely thermal RG-flows. This approach is
    applied to the ghost and gluon propagators as well as the
    ghost-gluon vertex in the Landau gauge. We present results in a
    temperature regime going from vanishing temperature to
    temperatures far above the confinement-deconfinement temperature
    $T_c$.  Our findings compare well with the lattice results
    available. }
\end{abstract}

\pacs{05.10.Cc, 11.15.Tk, 12.38.Aw, 11.10.Wx}

\maketitle

\pagestyle{plain}
\setcounter{page}{1}

\section{Introduction}\label{sec:introduction}

In the past two decades much progress has been made in our
understanding of the QCD phase diagram. This has been achieved in a
combination of first principle continuum computations in QCD, see
e.g.\
\cite{Pawlowski:2010ht,Binosi:2009qm,Fischer:2006ub,%
Alkofer:2000wg,Roberts:2000aa},
lattice simulations, see e.g.\
\cite{Philipsen:2011zx,Borsanyi:2010cj,Bazavov:2010sb}, as well as
computations in low energy effective models for QCD, see e.g.\
\cite{Fukushima:2010bq,Schaefer:2007pw,Megias:2004hj,%
Ratti:2005jh,Fukushima:2003fw,Dumitru:2003hp}. All
these methods have their specific strengths, and the respective
investigations are complementary.  In some applications, the methods
have even been applied in combination.

In recent years the first principle continuum approach as well as the
model computations have become far more quantitative. This is the more
interesting as at present they are one of the methods of choice for
finite density and non-equilibrium investigations that are necessary
for an understanding of heavy ion collisions. A full quantitative
understanding of the phase diagram of QCD from these methods requires
a good quantitative grip on the thermodynamics and finite temperature
dependence of Yang-Mills theory, for a review on thermal gluons see
\cite{Maas:2011se}, for applications of Dyson-Schwinger equations
(DSEs) for Yang-Mills propagators see
e.g. \cite{Gruter:2004bb,Cucchieri:2007ta}.\\
However, quantitatively reliable continuum results for correlation
functions in Yang-Mills theory at general temperatures are presently
not available. This concerns in particular the transition region of
the confinement-deconfinement phase transition, occurring at the
critical temperature $T_c$. The same applies to thermodynamic
quantities such as the pressure. Computations in Yang-Mills theory
done with hard-thermal loop or 2PI resummations do
not match the lattice results for temperatures $T \lesssim 3\,T_c$,
see e.g.\ \cite{Andersen:2011sf} and references therein. In turn, they are 
reliable for $T \gg T_c$. 

In the present work we put forward an approach with the functional
renormalisation group (FRG), for QCD-related reviews see
\cite{Litim:1998nf,Berges:2000ew,Pawlowski:2005xe,Gies:2006wv,%
  Schaefer:2006sr,Pawlowski:2010ht,Braun:2011pp,Rosten:2010vm}. The
FRG incorporates non-perturbative effects by a successive integration
of fluctuations related to a given momentum scale, also at
non-vanishing temperature, for applications in Yang-Mills theory see
e.g. \cite{Litim:1998nf,D'Attanasio:1996fy,Comelli:1997ru,Braun:2005uj,Litim:2006ag}.
It is applicable to all temperatures, in particular also for
temperatures $T \lesssim 3\, T_c$. The present implementation of the
FRG incorporates only thermal fluctuations, see e.g.\
\cite{Litim:2006ag}, and utilises the vacuum physics at vanishing
temperature as an input. This also allows us to study the effect of
global gauge fixing issues in the Landau gauge due to the Gribov
problem, for detailed discussions see
\cite{Fischer:2008uz,vonSmekal:2008ws}. A summary of the present work
and results is also presented in \cite{proceedings}.

In Section~\ref{sec:funflow} we give a brief introduction to the
FRG-approach to Yang-Mills theory. In particular, it is illustrated at 
the example of the two-point correlation functions, how flow
equations for correlation functions are derived from the flow of the
effective action.

Section~\ref{sec:local} is devoted to the important issue of locality of an
FRG-flow. We show how to minimise the momentum transfer present in the
diagram, and hence to minimise the sensitivity to the approximation at
hand. We also discuss the factorisation of diagrams at large external
momenta that is relevant for subleading momentum corrections. 

In Section~\ref{sec:thermalflow} we introduce the flow for thermal
fluctuations as that of the difference of flows at finite temperature
and vanishing temperature, see \cite{Litim:2006ag}. The locality
introduced in Section~\ref{sec:local} is used for improving the
stability of this difference, as well as having access to a trivial
initial condition for the thermal flow at large cut-off
scales. Variants of which have been very successfully used already for
the physics of scalar theories as well as fermion-boson mixtures, 
for reviews see \cite{Diehl:2009ma,Scherer:2010sv,Braun:2011pp}.

The approximation used in the present work is detailed in
Section~\ref{sec:approximation}, and some computational details are
discussed in Section~\ref{sec:comp}. Results for propagators and
vertices are presented in the Sections~\ref{sec:props+vertices}. They
are compared to lattice results, \cite{Maas:2011ez,Aouane:2011fv,
  Bornyakov:2010nc, Bornyakov:2011jm, Cucchieri:2000cy,%
  Cucchieri:2001tw,Cucchieri:2007ta, Cucchieri:2011di,
  Fischer:2010fx, Heller:1995qc, Heller:1997nqa, Maas:2011se}. Our
observations are shortly summarised in Section~\ref{sec:summary}.

\section{Yang-Mills correlation functions from
  RG-flows}\label{sec:funflow}
 
In this section we introduce the functional RG setting for Yang-Mills
theories which we use for the computation of propagators and
vertices. Central to this approach is the scale-dependent effective
action $\Gamma_k$, where quantum fluctuations with momenta $p^2\gtrsim
k^2$ are already integrated out, and $k$ denotes an infrared cut-off
scale. By varying the cut-off scale $k$ one thereby interpolates
between the classical action in the ultraviolet and the full quantum
effective action in the infrared, where the cut-off is removed. An
infinitesimal change of $k$ is described by a flow equation for
$\Gamma_k$, and the interpolation can be achieved by integrating this
flow. \smallstep

Within the standard Faddeev-Popov gauge fixing procedure in the Landau
gauge,
\begin{equation} 
  \pd_\mu A_\mu^a=0\,,\qquad a=1,...,N_c^2-1,\label{eq:landau}
\end{equation}
the classical gauge fixed action for $SU(N_c)$ Yang-Mills theory
in Euclidean space-time is given by
\begin{eqnarray} 
S=\int_x\left(\014 F_\mn^a F_\mn^a+\bar c^a\pd_\mu D_\mu^{ab} c^b 
\right)\,,\label{eq:fixedaction} 
\end{eqnarray} 
with $\int_x=\int d^4 x$. In \eq{eq:fixedaction} we have already
introduced the ghost action with the ghost fields $\bar c^a$ and
$c^a$. The field strength $F$ and the covariant derivative $D$ are
given by
\begin{eqnarray}
  F^a_\mn&=&\pd_\mu A^a_\nu-\pd_\nu A_\mu^a+gf^{abc} A_\mu^b 
  A_\nu^c\,,\nn\\
  D_\mu^{ab}&=&\delta^{ab}\pd_\mu+g f^{acb} A_\mu^c\,.\label{eq:defFD}\end{eqnarray} 
The gauge fields are transveral due to the Landau gauge \eq{eq:landau}. Hence operators 
and the inverse of 
operators are defined on the transversal subspace. The suppression of the 
infrared fluctuations is achieved via a modification of the
propagator with $S \to S+\Delta S_k$ and
\begin{eqnarray}\label{eq:dSk}
\Delta S_k=
\012 \int_p \, A^{a}_\mu\, R_{\mu\nu}^{ab} \, A^{b}_\nu  
+\int_p \, \bar c^{a}\, R^{ab}\, c^{b}\,, 
\end{eqnarray} 
with $\int_p=\int d^4 p/(2\pi)^4$. 
\begin{figure}[t]
\includegraphics[width=.8\columnwidth]{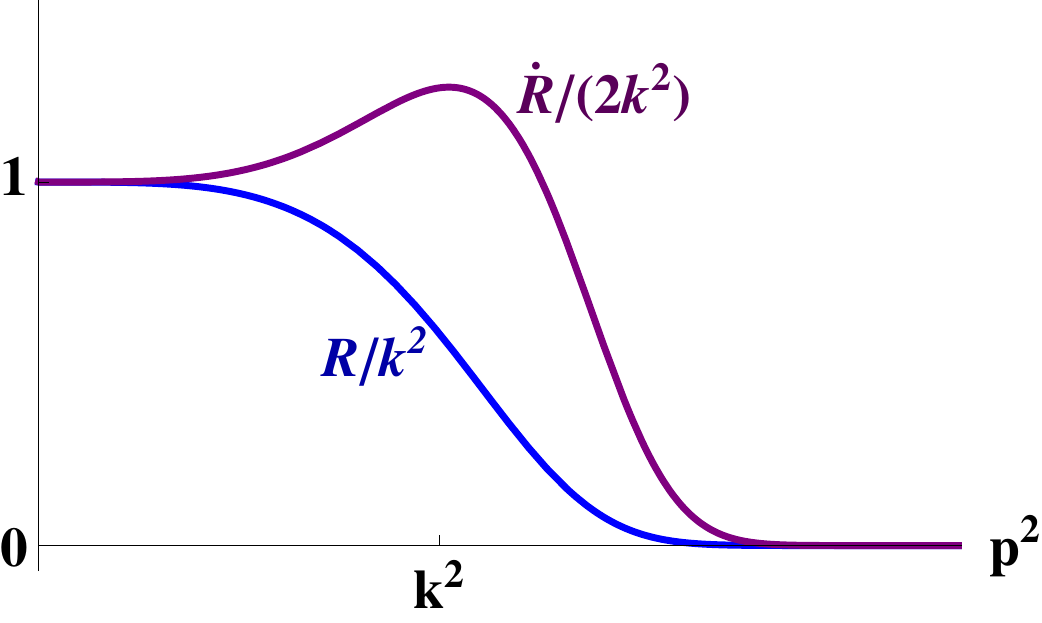}
\caption{Regulator $R(p^2)$ and its (logarithmic) cut-off scale
  derivative $\partial_t R(p^2)=\dot R(p^2)$. }
\label{fig:R}
\end{figure}
The momentum-dependent regulator
functions $ R_{\mu\nu}^{ab}$ and $R^{ab}$ implement the infrared
cut-off at the momentum scale $p^2\approx k^2$ for gluon and ghost
respectively. They carry a Lorentz and gauge group tensor structure
and are proportional to a dimensionless shape function $r$,
schematically we have $R(p^2)\propto p^2 \,r(p^2/k^2)$. An example for
$R(p^2)$ and its logarithmic scale derivative is given in
Fig.~\ref{fig:R}. As we work in Landau gauge, \eq{eq:landau}, we can
restrict ourselves to a transversal gluon regulator, $p_{\mu}
R_{\mu\nu}^{ab}(p)=0$.

\begin{figure}[t]
\includegraphics[width=.8\columnwidth]{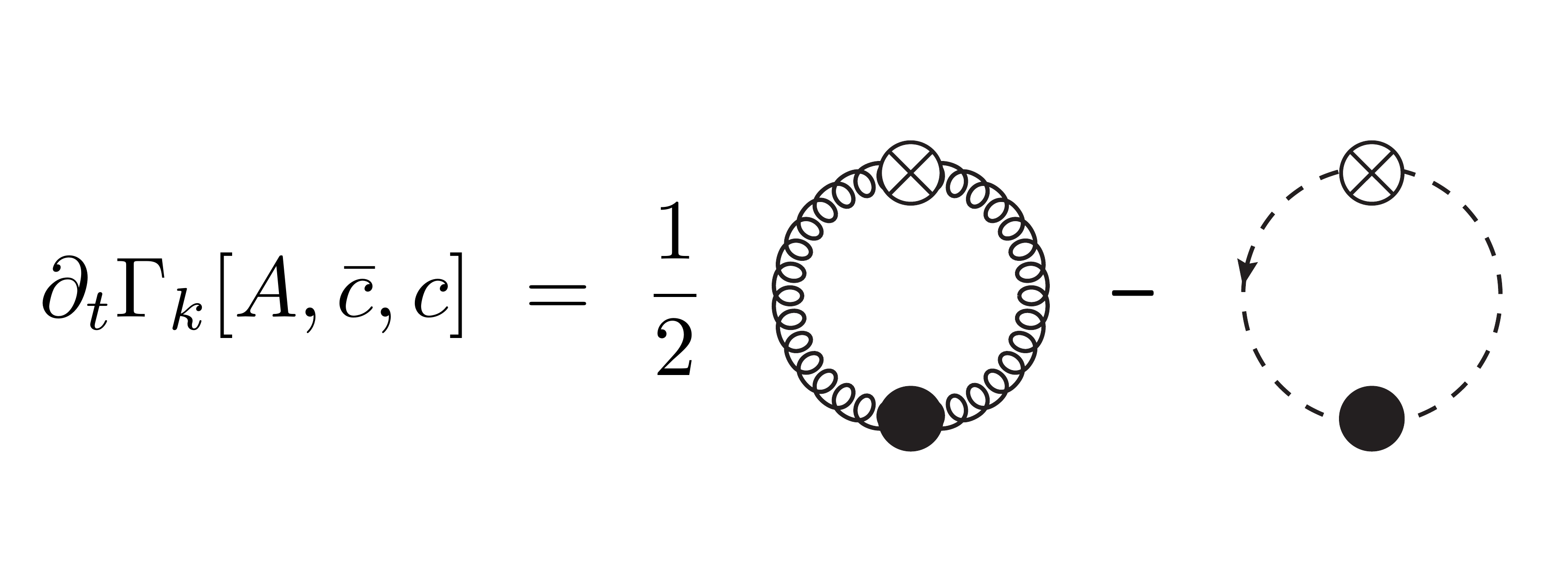}
\caption{Functional flow for the effective 
action. Lines with filled circles denote fully dressed field dependent propagators 
\eq{eq:G}. Crossed circles denote the regulator insertion $\partial_t R_k$. }
\label{fig:funflow}
\end{figure}

With the notation $\varphi=(A,c,\bar c)$ we write the flow equation
for the Yang-Mills effective action $\Gamma_k[A,c,\bar c]$ at finite
temperature $T$ as
\begin{eqnarray} \nonumber 
  \partial_t \Gamma_{k}[\varphi] &= & 
  \frac{1}{2} \sumint_p\ G^{ab}_{\mu\nu}[\varphi](p,p)
  \, {\partial_t} R_{\nu\mu}^{ba}(p)\\ &&-
  \sumint_p\ G^{ab}[\varphi](p,p)
  \, {\partial_t} R^{ba}(p)\,,
\label{eq:funflow}\end{eqnarray} 
where $t=\ln k$, and the momentum integration measure at finite
temperature is given by
\begin{eqnarray}\label{eq:matsubaras}
\sumint_p=T \sum_{n\in \mathbb{Z}}\int \0{d^3 p}{(2 \pi)^3}\,,
\quad {\rm with}\quad  p_0=2 \pi T n\,,
\end{eqnarray} 
where the integration over $p_0$ turns into a sum over Matsubara
frequencies. Both, gluons and ghosts have periodic boundary conditions,
$\varphi(x_0+1/T,\vec x)=\varphi(x_0,\vec x)$, which is reflected
in the Matsubara modes $2 \pi T n$ with a thermal zero mode for $n=0$. 
At vanishing temperature we have $\sum\!\!\!\!\!\!\int\to
\int_p$. The full field dependent propagator for a propagation from
$\varphi_1$ to $\varphi_2$ is given by
\begin{eqnarray}\label{eq:G}
  G_{\varphi_1\varphi_2}[\varphi](p,q)=\left(\0{1}{\Gamma_{T,k}^{(2)}[\varphi]
      +R_k}\right)_{\varphi_1\varphi_2}(p,q)\,.
\end{eqnarray} 
In \eq{eq:G} we also used the regulator function in field space,
$R_{k,\varphi_1\varphi_2}$ with
\begin{eqnarray}\label{eq:Reg}
R_{k,A_\mu^a A_\nu^b}= R_{\mu\nu}^{ab}\,,  \qquad  R_{k,\bar c^a
  c^b}=-R_{k, c^b
  \bar c^a}=R^{ab}\,. 
\end{eqnarray} 
The flow is finite in both, the infrared and the
ultraviolet, by construction.  Effectively, the momentum integration
in \eq{eq:funflow} only receives contributions for momenta in the
vicinity of $p^2\lesssim k^2$. Consequently, it has a remarkable
numerical stability.  The flow solely depends on dressed vertices and
propagators, leading to consistent RG-scaling on either side of
\eq{eq:funflow}. Diagrammatically, the flow in \eq{eq:funflow} is depicted in
Fig.~\ref{fig:funflow}.

The structure of the flow \eq{eq:funflow}, or 
Fig.~\ref{fig:funflow}, is more apparent in a formulation with the
fields $\varphi$.  Written in these fields the cut-off term reads $\Delta
S_k[\varphi]= \s012 \varphi\cdot R_k\cdot \varphi$, where the dot
indicates the contraction over species of fields including space-time
or momenta, Lorentz and gauge group indices.  Note that the cut-off
term now includes two times the ghost term which cancels the global
factor $1/2$ and leads to \eq{eq:dSk}.  In this short hand notation
the flow equation takes the simple and concise form
\begin{eqnarray}\label{eq:funflowstruc}
\partial_t \Gamma_{k}[\varphi] &= & 
  \frac{1}{2} \Tr\, G[\varphi](p,p)
  \cdot {\partial_t} R_k(p)\,,
\end{eqnarray}
where the trace includes a relative minus sign for the
ghosts. \Eq{eq:funflowstruc} is well-suited for structural
considerations.

\begin{figure}[t]
\includegraphics[width=6.5cm]{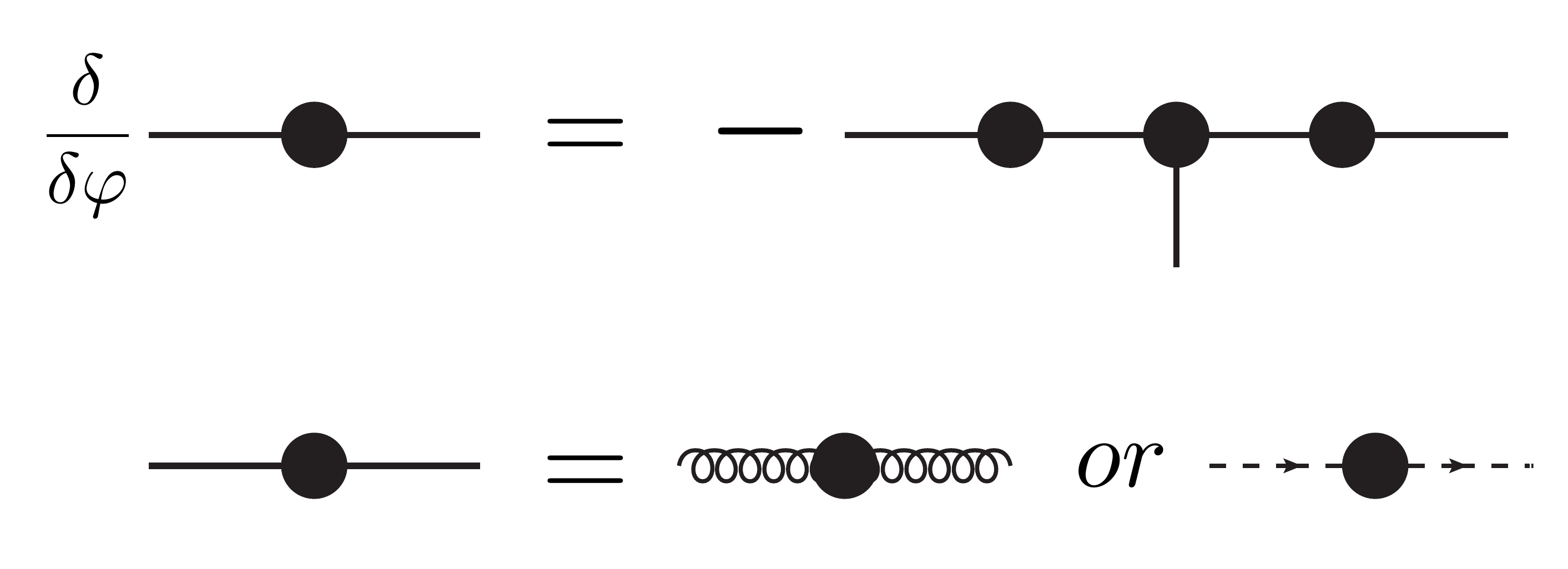}
\caption{Field derivative of the propagator. The solid line stands for
  either a gluon or a ghost propagator and the vertex is depicted by a filled circle. }
\label{fig:derG}
\end{figure}

\begin{figure}[t]
\includegraphics[width=\columnwidth]{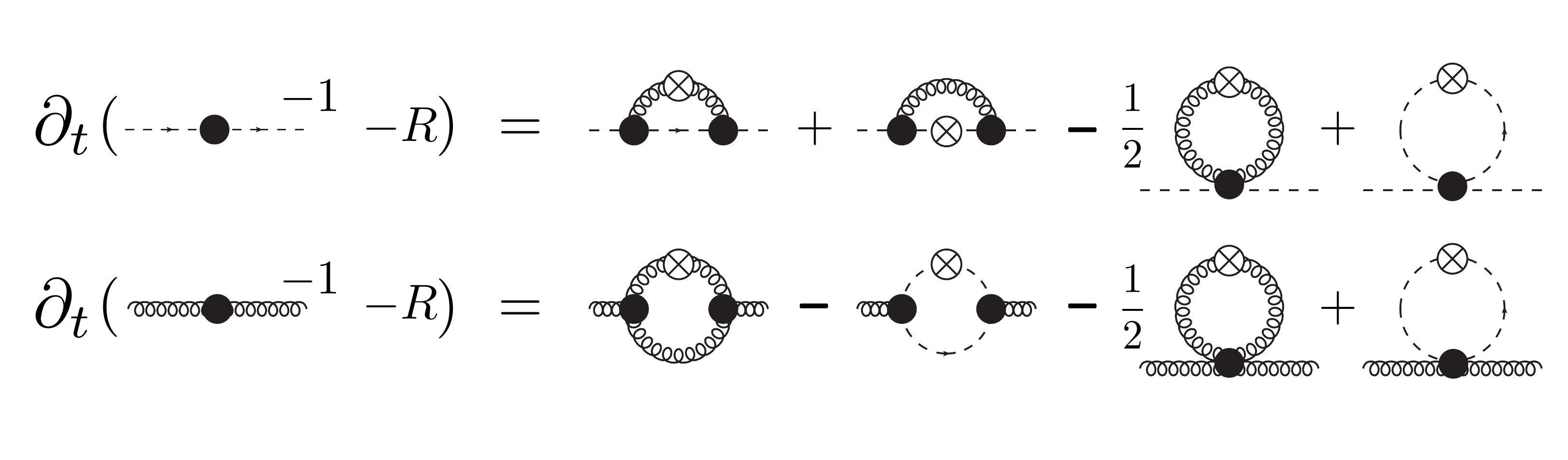}
\caption{Flow equations for the ghost and gluon propagators.
  Vertices with filled circles denote fully dressed vertices. All
  propagators are fully dressed, the filled circles for the internal
  ones have been omitted for the sake of clarity.  Crossed circles
  denote the regulator insertion $\partial_t R_k$.}
\label{fig:YM_props}
\end{figure}

Flow equations for propagators and vertices are obtained from
\eq{eq:funflow} or \eq{eq:funflowstruc} by taking derivatives with respect to 
$\varphi$. The computation of $\varphi$-derivatives on the rhs of 
\eq{eq:funflowstruc} only requires the equation for the field
derivative of the propagator and the notion of the full one-particle
irreducible (1PI) $n$-point functions. The latter are defined by
\begin{eqnarray}\label{eq:Gamman}
  \Gamma_{k}^{(n)}(p_1,...,p_n)=\0{\delta\Gamma_{k}}{\delta 
    \varphi(p_1)\cdots\varphi(p_n)}\,.
\end{eqnarray}
The field derivative of the propagator is given by  
\begin{eqnarray}\label{eq:derG}
  \0{\delta}{\delta \varphi} G[\varphi]= 
  - G[\varphi]\cdot \Gamma_{k}^{(3)}[\varphi]\cdot G[\varphi]\,,
\end{eqnarray}
and diagrammatically depicted in Fig.~\ref{fig:derG}. 
With the above equations one readily derives the flow of 1PI $n$-point
functions.  Again, this can be nicely illustrated diagrammatically at
the example of the propagators, which are the key objects in the
approach. We take two field derivatives of the flow equation. Applying
this to its diagrammatical form displayed in Fig.~\ref{fig:funflow} we
are led to the flow equations given in Fig.~\ref{fig:YM_props}.
The flow equations only contain one loop diagrams in the full
propagators, which is a consequence of the exact one-loop form of the
flow for the effective action, see \eq{eq:funflow} and
Fig.~\ref{fig:funflow}. Moreover, they only depend on the full
vertices \eq{eq:Gamman}. Analogously to the flow of the two-point
functions, the flows of $n$-point functions only contain one-loop
diagrams that depend on full propagators and vertices. 

The propagators and vertices are purely transversal due to the
transversality of the gauge field. The purely transversal correlation
functions vanish by contracting one of the Lorentz indices with its
momentum, $(p_i)_{\mu_i} \Gamma^{(n)\,T}_{\mu_1\cdots
  \mu_i\cdots\mu_m}=0$ for $i=1,...,m$. Note that in general $n\neq m$
due to the ghosts. Since the propagators are purely transversal in the
Landau gauge, the purely transversal correlation functions in the
Landau gauge form a closed system of flow equations: the flow of a
purely transversal correlation function only depends on purely
transversal correlation functions and carry the whole dynamics. Any
observables can be built-up from the purely transversal correlation
functions.

In turn, the flow of correlation functions with at least one
longitudinal direction, $\Gamma^{(n)\,L}$, depends on both,
$\Gamma^{(n)\,T}$ and $\Gamma^{(n)\,L}$. Moreover, they also obey
modified Slavnov-Taylor identities, see
\cite{Ellwanger:1994iz,Freire:2000mn,Pawlowski:2005xe,Gies:2006wv} and
references therein.This situation is summarised in
\begin{eqnarray}\nonumber 
\partial_t  \Gamma^{(n)\,T}&=& {\rm Flow}_n^T[\{\Gamma^{(m)\,T}\}]\,,
\\[1ex]\nonumber 
\partial_t  \Gamma^{(n)\,L}&=& {\rm Flow}_n^L[\{\Gamma^{(m)\,T}\,,
\Gamma^{(m)\,L}\}]\,,\\[1ex]
 (p)_{\mu} \Gamma^{(n)\,L}_{\mu\mu_2\cdots\mu_m}&=& {\rm mSTI}_n[\{\Gamma^{(m)\,T},
\Gamma^{(m)\,L}\}]\,, 
\label{eq:purelyTflow+Lflow}\end{eqnarray}
where $m\leq n+2$.  The modified Slavnov-Taylor identities converge to
the standard ones for vanishing cut-off scale $k=0$. The above system
comprises the full information about the correlation functions in the
Landau gauge.  Interestingly, the hierarchy of flow equations for the
purely transversal correlation functions can be solved independently
and carries the full dynamics of Yang-Mills theory. In this context we
remark that vertex constructions aided by Slavnov-Taylor identities
implicitly utilise an assumed uniformity of correlation functions, that is 
\begin{eqnarray}\label{eq:differentiable} 
  \partial_{p_\mu}  \Gamma^{ (n) } < \infty \quad {\rm for\ all\ } (p_1,...,p_n)\,.  
\end{eqnarray} 
This works well in perturbation theory but has to be taken with a grain
of salt in the non-perturbative regime. 

It is also worth emphasising that \eq{eq:purelyTflow+Lflow} does
not depend on the way how the Landau gauge is introduced. The Landau
gauge can be also represented as the limit of covariant gauges with
the gauge action $1/(2 \xi) \int_x (\partial_\mu A^a_\mu)^2$ and gauge
fixing parameter $\xi$. Here $\xi=0$ signals the Landau gauge. In this
case in general one also introduces a regularisation for the gauge
mode, schematically given by
\begin{equation}
 R^{\rm L} = \lim_{\xi\to 0}\0{1}{\xi} \Pi^L\, p^2 r(p^2/k^2)\,, 
\end{equation}
with the projection operators 
\begin{eqnarray}
  \Pi^T_{\mu \nu}(p)&=&\delta_{\mu \nu}-p_{\mu}p_{\nu}/p^2\,,\nn\\
  \Pi^L_{\mu \nu}(p)&=&p_{\mu}p_{\nu}/p^2 \,.
  \label{eq:projections0} \end{eqnarray}
on transversal and longitudinal degrees of freedom, respectively. 
Still, the longitudinal mode does not play any r$\hat{\rm
  o}$le for the flow of correlation functions as $G \cdot \partial_t
R^{\rm gauge}\cdot G\to 0$ for $\xi\to 0$ and $\lim_{\xi\to 0} G$ is
purely transversal. Note however, that $1/2\,\Tr\, R^{\rm
  gauge}\cdot G$ does not vanish. Upon $t$-integration it gives the
thermal pressure for the gauge mode, namely the Stefan-Boltzmann
pressure of $N_c^2-1$ fields, see \cite{Litim:2006ag}.

\section{Locality of RG-flows}\label{sec:local}
\begin{figure}[t]
\includegraphics[width=\columnwidth]{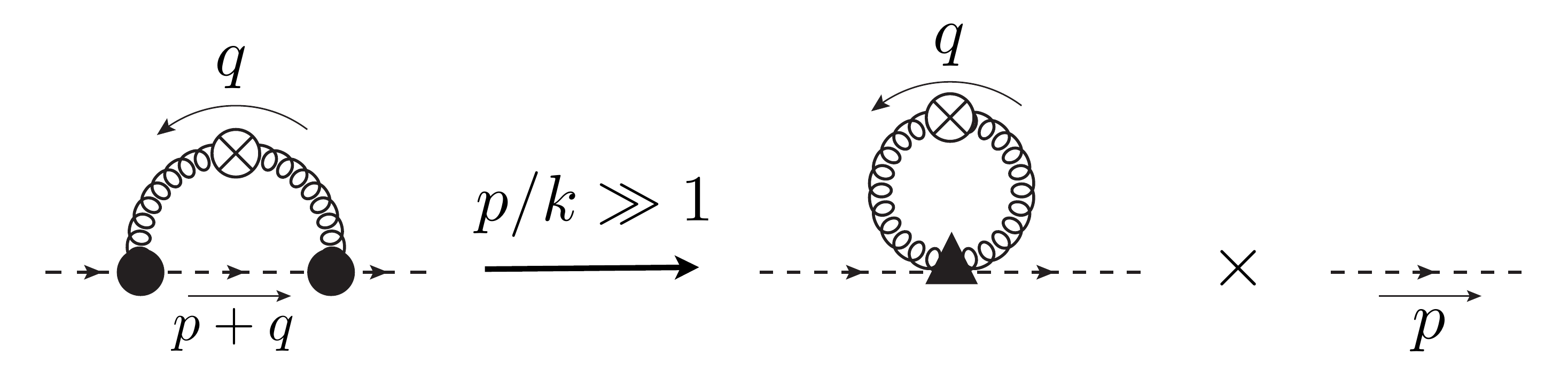}
\caption{Factorisation in leading order for large momenta for the
  first diagram (cutted gluon line) in the flow of the (inverse) ghost
  propagator in Fig.~\ref{fig:YM_props} . The triangle stands for the
  product of the two vertices at $q=0$ and reads $-p_\mu p_{\mu'} f^{a
    cd} f^{b c' d'}$. }
 \label{fig:factorisation}
\end{figure}

The loops on the rhs of Fig.~\ref{fig:YM_props} only received
contributions from momentum fluctuations with $p^2 \lesssim k^2$ due
to the regulator insertion $\partial_t R$, see
Fig.~\ref{fig:R}. However, the external momenta are not limited by
such a constraint. Indeed, for large external momenta, $p^2/k^2 \gg 1$
the flow factorises at leading order: the tadpole diagram with the
four-gluon vertex tends to a constant. The other tadpole diagrams
vanish with powers $\lesssim k^2/p^2$. The related four-point
functions are not present on the classical level and hence decay at
least with $k^2/p^2$. This intuitive statement can be proven easily
with the help of the respective flows and the factorisation present
there. Here, we explain the factorisation at the relevant example of
the three-point function diagrams, see also
Fig.~\ref{fig:factorisation} for the example of the flow of the ghost
propagator. These diagrams factorise for large external momenta, one
factor being the uncutted internal line evaluated at $p^2$. For
$p^2/k^2 \to\infty$ we have,
\begin{eqnarray}\nonumber 
  && \hspace{-.7cm}\Tr\,\left[ \left(G \dot R G\right)(q)\cdot 
    \Gamma^{(3)}(q,p+q) 
    \cdot G(p+q) \cdot  \Gamma^{(3)}(q+p,q)\right] \\\nonumber 
  & \to&   \Tr\,\left[\left(G \dot R G\right)(q)\cdot \Gamma^{(3)}(0,p) 
    \cdot G(p)  \cdot \Gamma^{(3)}(p,0)\right]\\ 
  && +\,{\rm higher\ order\ terms} \,, 
\label{eq:factorisation}\end{eqnarray}
where the trace also integrates over loop momenta $q$. 
In \eq{eq:factorisation} we have assumed that all momentum components
are suppressed by the regulator: $R$ is a function of $p^2$ and not
e.g. of solely spatial momentum squared, $\vec p^{\,2}$.  Note also that
for kinetic and/or symmetry reasons the leading order in line two of
\eq{eq:factorisation} may vanish. This even supports the
factorisation. The interchange of integration and limit and hence the
factorisation works as the diagram is still finite with the uncutted
line being removed.

The general factorisation in \eq{eq:factorisation} can be nicely
illustrated diagrammatically at the example of the flow of the ghost
propagator Fig.~\ref{fig:YM_props}. In Fig.~\ref{fig:factorisation} we
display the factorisation of the first diagram in
Fig.~\ref{fig:YM_props}. 
The terms in the second line of \eq{eq:factorisation} or on the rhs of
Fig.~\ref{fig:factorisation} are already subleading as the flow
usually is peaked at momentum scales $p^2\lesssim k^2$.  However, they
are already quantitatively relevant at vanishing temperature but turn
out to be crucial for the correct thermodynamics, in particular for
the slow approach to the Stefan-Boltzmann limit for large
temperatures. Potentially, they also play an important r$\hat{\rm
  o}$le for the thermodynamics in non-relativistic systems, where they
supposedly relate to the Tan-relations
\cite{Tan:2008-1,Tan:2008-2,Tan:2008-3} in the context of many-body
physics; for FRG-reviews see e.g.\ \cite{Diehl:2009ma,Scherer:2010sv}.

Still we have to reconcile the above polynomial decay with external
momenta $p$ with the well-known exponential decay of thermal
fluctuations with the standard suppression factor $\exp(-m/T)$ in the
presence of a mass scale $m$. In the present case, this mass scale can
be either the cut-off scale $k$ or the physical mass scale of
Yang-Mills theory, $\Lambda_{\rm QCD}$, which is directly linked to the
critical temperature $T_c$ of the confinement-deconfinement phase
transition. The exponential thermal damping factor originates from the
full Matsubara sum, and is strictly not present if only the lowest
Matsubara frequencies are taken into account. A four-dimensional
regulator depending on four-dimensional loop momentum $q^2=(2 \pi T
n)^2+\vec q^2$ cuts the Matsubara sum, and hence only leads to the
polynomial decay of the flow.  The exponential suppression is then
built-up successively with the flow. The above properties can be
already very clearly seen and understood at the example of
perturbative one-loop flows.

The full Matsubara sum is reintroduced for regulators only depending
on $\vec p^{\,2}$.  They are frequently used in finite temperature
applications of the FRG as they allow for an analytic summation of the
Matsubara sums if only the trivial frequency dependence is taken into
account, see e.g.\
\cite{Litim:1998nf,Litim:2006ag,Blaizot:2006rj,Braun:2009si}. However,
in the present work we consider full propagators and vertices, that is
with non-trivial frequency and momentum dependence. Moreover, as
lower-dimensional regulators introduce an additional momentum or
frequency transfers in the flow we therefore refrain from using them
here.

Apparently, the large momentum contributions also weaken the locality
of the flow present in the loop momenta as they induce a momentum
transfer: the flows at a given cut-off scale $k$ carry physics
information about larger momentum scales. In turn this entails that
any local approximation does not fully cover this momentum transfer.
Now we utilise the freedom of reparameterising the effective action as
well as the choice of the regulator in order to reinstall the locality
of the flow for the two-point functions $\Gamma_{k}^{(2)}$. This
minimises the systematic error of a given approximation, see
\cite{Pawlowski:2005xe}. To that end, we rewrite the regulator term 
in \eq{eq:dSk} as follows, 
\begin{eqnarray}\label{eq:dSkhat}
\Delta S_k=
\012 \int_p \, A_{k,a}^\mu\, \hat R_{\mu\nu}^{ab} \, A_{k,b}^\nu  
+\int_p \, \bar c_{k,a}\, \hat R^{ab}\, c_{k,b}\,.
\end{eqnarray} 
The fields $\phi=(A_k, c_k,\bar c_k)$ in \eq{eq:dSkhat} relate to the
cut-off independent fields $\varphi=(A,c,\bar c)$ in the classical
action via a cut-off and momentum dependent rescaling,
\begin{eqnarray}\label{eq:rescaling} 
\phi(p)= \hat Z^{1/2}_{\phi,k}(p)\varphi(p)\,\quad {\rm with}\quad 
 \partial_t \phi(p)=\hat\gamma_\phi(p)\, \phi(p)\,, 
\end{eqnarray} 
where the derivative is taken at fixed $\varphi$. We will use the
natural definition $\hat\gamma_{\bar
  c}=\hat\gamma_c$. \Eq{eq:rescaling} implies
\begin{eqnarray}\label{eq:gamma+eta}
  \hat\gamma_\phi(p)=\partial_t \log \hat Z_\phi(p)\,,
  \quad {\rm and}\quad 
  R=\hat Z \cdot \hat R\,.
\end{eqnarray}
The field reparametrisation in \eq{eq:rescaling} does not change the
effective action, in particular the regulator term $\Delta S_k$ does
not change. It simply amounts to rewriting the effective action in
terms of the new fields,
\begin{eqnarray}\label{eq:hatGamma}
  \hat\Gamma_{k}[\phi]=\Gamma_k[\varphi]\,.
\end{eqnarray}
Then, $\phi$-derivatives $\hat\Gamma^{(n)}_k$ of the
effective action $\Gamma_k=\hat\Gamma_k$ are given by \eq{eq:Gamman} with
$\varphi\to \phi$,
\begin{eqnarray}\label{eq:hatGamman}
  \hat\Gamma_{k}^{(n)}(p_1,...,p_n)=\0{\delta\hat\Gamma_{k}}{\delta 
    \phi(p_1)\cdots\delta\phi(p_n)}\,.
\end{eqnarray}
As the fields $\phi$ and $\varphi$ only differ by a momentum dependent
rescaling with $\hat Z^{1/2}$, the correlation functions are related by a
simple rescaling with powers of $\hat Z^{1/2}$, 
\begin{eqnarray}
  \Gamma^{(n)}(p_1,...,p_n) &=& \prod_{i=1}^{n} \hat Z^{1/2}_{\phi_i}(p_i) 
  \, \hat\Gamma^{(n)}(p_1,...,p_n)
  \,.
\label{eq:hatunhat}\end{eqnarray}
The rescaling with $\hat Z$ is up to our disposal, and we shall use it
to minimise the momentum transfer in the flow equation of the
two-point function by eliminating the subleading terms in the flow
exemplified in \eq{eq:factorisation}. Note that this does not remove
the related contributions, the momentum transfer is still present but
does not feed-back directly in the flow, see
\cite{Pawlowski:2005xe}. This is also elucidated below.

The flow equation for the effective action now receives further
contributions from the $k$-dependence of the fields $\phi$.  We are
finally led to the following flow for
$\hat\Gamma_{k,T}[\phi]=\Gamma_{k}[\varphi]$, see
\cite{Pawlowski:2005xe},
\begin{eqnarray}\nonumber 
&&\hspace{-1.7cm}\left({\partial_t}+\sum_i 
\sumint_p \hat\gamma_{\phi_i}(p) \phi_i(p)
\0{\delta}{\delta\phi_i(p)}\right)
\hat\Gamma_{T,k}[\phi] =   \\ \nonumber   
&& \frac{1}{2} \sumint_p \ \hat G^{ab}_{\mu\nu}[\phi](p,p)
\, ({\partial_t}+2 \hat\gamma_A(p)) \hat R_{\nu\mu}^{ba}(p)\\  
&&-
\sumint_p \hat G^{ab}[\phi](p,p)
\, ({\partial_t}+2 \hat\gamma_C(p)) \hat R^{ba}(p)\,,
\label{eq:hatfunflow}\end{eqnarray} 
where $\hat G[\phi](p,q)= \bigl(\hat\Gamma_{k}^{(2)}[\phi]+\hat R
\bigr)^{-1}(p,q) $ denotes the full regularised propagator for the
propagation of $\phi$, see \eq{eq:G} with $\varphi\to \phi$ and
$\Gamma^{(2)}_k \to \hat \Gamma^{(2)}_k$. The functional flow in
\eq{eq:hatfunflow} looks rather complicated, but it is simply a
reparameterisation of the standard flow in \eq{eq:funflow}, or
\eq{eq:funflowstruc}. In the condensed notation introduced in
Section~\ref{sec:funflow} this is more apparent, the flow equation
\eq{eq:hatfunflow} then reads
\begin{equation}\label{eq:hatfunflowstruc}
\left({\partial_t}+\phi\cdot\hat\gamma_{\phi}\cdot \0{\delta}{\delta\phi}\right)
  \hat\Gamma_{T,k}[\phi]= \frac{1}{2} \Tr \, \hat G[\phi] \cdot 
  \, ({\partial_t}+2 \hat\gamma_\phi)\cdot \hat R_k\,. 
\end{equation}
\Eq{eq:hatfunflowstruc} illustrates that we have only reparametrised
the fields in a scale-dependent way. \smallstep

Taking two derivatives with respect to \ $\phi_1(p)$ and $\phi_2(q)$ of
\eq{eq:hatfunflow} at vanishing ghost fields and constant gauge
field we schematically get the flow
\begin{equation}
  \left(\partial_t+ \hat \eta_{\phi_1}(p)\,\right)
\hat\Gamma_{\phi_1\phi_2}^{(2)}(p) =   {\rm Flow}^{(2)}_{\phi_1\phi_2}(p)\, ,
\label{eq:2pointflow}\end{equation} 
where the rhs of \eq{eq:2pointflow} stands for the $\phi_1(p)$ and
$\phi_2(q)$ derivative of the rhs of \eq{eq:hatfunflow}, 
\begin{equation}\label{eq:defofflow}
 {\rm Flow}^{(2)}_{\phi_1\phi_2} = \0{\delta^2}{\delta\phi_1\delta\phi_2} \left(
\frac{1}{2} \Tr \, \hat G[\phi] \cdot 
  \, ({\partial_t}+\hat\eta_\phi)\cdot  \hat R_k\right)\,, 
\end{equation}
and 
\begin{eqnarray}\label{eq:hateta}
\hat\eta_\phi(p)=2 \hat\gamma_\phi(p)=\partial_t \log \hat Z_k(p)\,,
\end{eqnarray} 
is the `anomalous' dimension of the propagator related to the
rescaling of the fields with $\hat Z$. In \eq{eq:2pointflow} we have
used that for vanishing ghosts the two-point functions are
diagonal/symplectic in field space. The only non-vanishing components
are $\hat\Gamma_{k,AA}^{(2)}$ and $\hat\Gamma_{k,c\bar c}^{(2)}=
-\hat\Gamma_{k,\bar c c}^{(2)}$. The two-point functions for constant
gauge fields are also diagonal in momentum space, that is
\begin{eqnarray}\nonumber
\hat\Gamma_{k}^{(2)}(p,q) &=&
\hat\Gamma_{k}^{(2)}(p)(2 \pi)^4 \delta(p-q)\,,\\
{\rm Flow}^{(2)}(p,q) &=& 
{\rm Flow}^{(2)}(p) (2 \pi)^4\delta(p-q)\,,
\label{eq:diagonal}\end{eqnarray}
and the relation \eq{eq:hatunhat} for the two-point functions reads 
\begin{equation} \label{eq:paraGform}
\Gamma^{(2)}(p)=\hat Z(p)\,\hat \Gamma^{(2)}(p)\,.
\end{equation} 
Now we can come back to the question of locality of the flow. Our aim
is to remove the large momentum tail displayed in
\eq{eq:factorisation} from the flow in order to keep its locality in
momentum space, i.e. to minimise the momentum transfer. This is
achieved by demanding
\begin{eqnarray} \label{eq:locality} 
\left.\partial_t \hat\Gamma_{k}^{(2)}(p)\right|_{p^2>(\lambda k)^2}\equiv 0 \,,
\end{eqnarray} 
which implies 
\begin{eqnarray} 
\hat\eta_\phi(p)&=&
\0{{\rm Flow}^{(2)}(p)}{\hat\Gamma_{k}^{(2)}(p)}
\theta(p^2 -(\lambda k)^2)\,. \label{eq:hatetaspecific}
\end{eqnarray} 
\Eq{eq:locality} entails that the momentum transfer is switched off
for momenta larger than the cut-off scale. The factor $\lambda$
controls this scale and can be used for an error estimate. 

After having computed the localised correlation functions
$\hat\Gamma^{(n)}$, we can derive the correlation functions
$\Gamma^{(n)}$ via rescaling with powers of $\hat Z$, see
\eq{eq:hatGamman},\eq{eq:paraGform}. The scaling factor is computed by
integrating $\hat \eta_k$ defined in \eq{eq:hateta},
\begin{eqnarray}
\hat{Z}_k(p;T) &=& \hat{Z}_{k=0}(p;T) \ {\rm exp}\!\left\{ 
\int_{0}^{k}dt' \ \hat{\eta}_{k'} (p;T)\right\}\,.
\label{eq:hatZint}\end{eqnarray}
This determines $\hat Z_k$ up to a $k$-independent function, and we use 
\begin{eqnarray}\label{eq:hatZ0}
\hat Z_{\phi,k=0}(p;T=0)=1\,. 
\end{eqnarray} 
For the choice \eq{eq:hatZ0} the two sets of correlation functions
agree in the vacuum at $k=0$. 

As the flow of general correlation functions can be written down
solely in terms of $\hat \Gamma^{(n)}$, the relation \eq{eq:locality}
with \eq{eq:hatetaspecific} eliminates the momentum transfer (in
$\Gamma^{(2)}$) from the flow. Note however, that a remnant of it is
still present via the factor $2 \hat\gamma_\phi=\hat\eta_\phi$ on the
rhs of the flow \eq{eq:hatfunflowstruc}. For regulators that decay
sufficiently fast for momenta $p^2 \gg k^2$ this is quantitatively
negligible. Indeed, for regulators which vanish identically for
momenta bigger than $\lambda k$ the momentum transfer now is described
solely by $\hat \eta_\phi(p)$ and decouples completely. We also remark
in this context, that the above construction and the definition
\eq{eq:hatZ0} leading to \eq{eq:locality} with $\lambda=1$ can be
deduced by evoking functional optimisation for momentum-dependent
approximations, see \cite{Pawlowski:2005xe}. The above heuristic
arguments entail that optimisation restores the locality of the flow
also in general momentum- and frequency-dependent approximations.

\section{Thermal fluctuations}\label{sec:thermalflow}
\begin{figure}[t]
\includegraphics[width=.8\columnwidth]{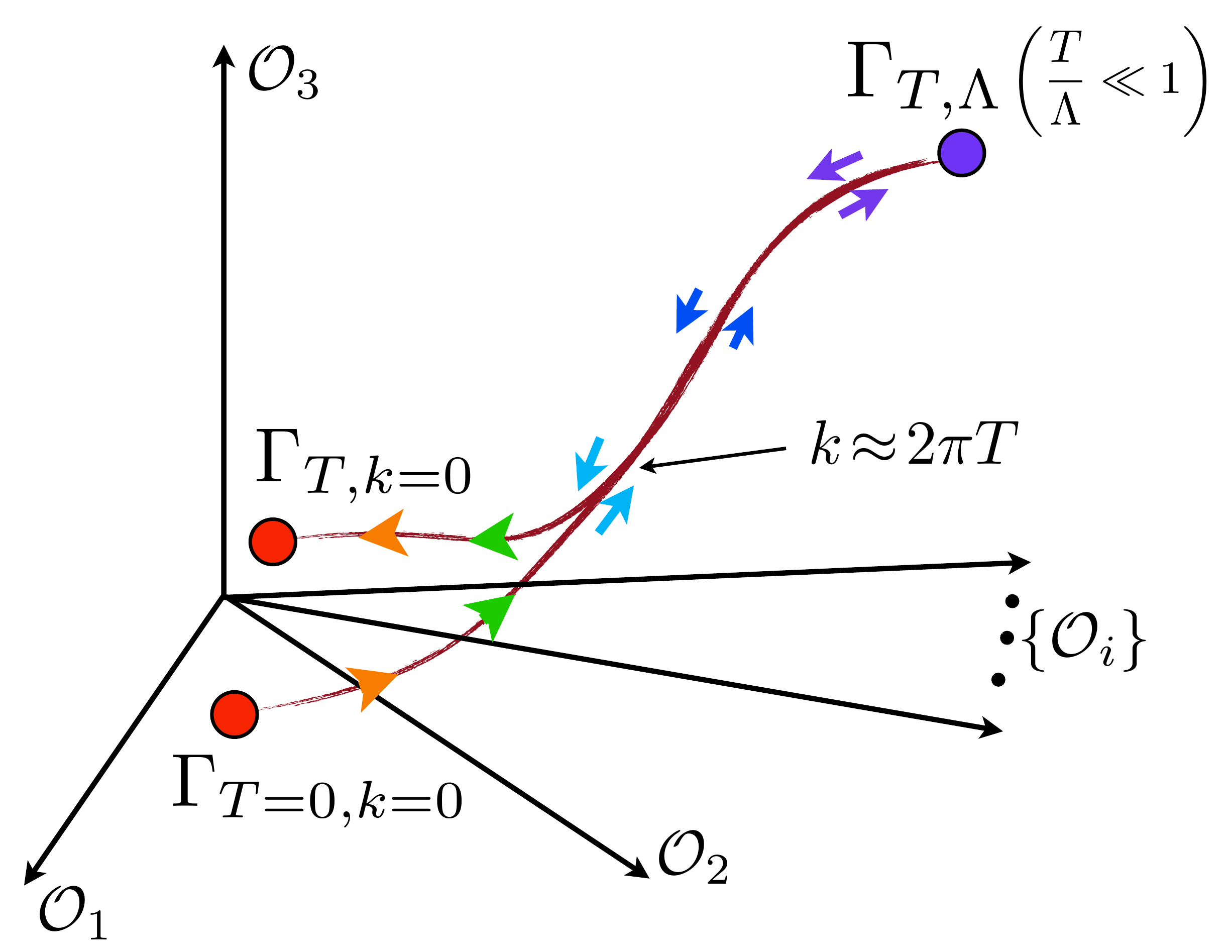}
\caption{Flow at vanishing temperature from $k=0$ to $k=\Lambda$, and
  flow at $T\neq 0$ from $k=\Lambda$ to $k=0$ with $\Lambda/T\gg
  1$. The flow is described in theory space, and the axes label
  (orthogonal) couplings/observables $\CO_i$ which serve as expansion
  coefficients of the effective action,
  e.g. $\CO_1=\Gamma^{(2)}(p=0)$. The flows start to deviate at
  $k\approx T$, see the discussion below.}
 \label{fig:thermalflow}
\end{figure}

The flow equation \eq{eq:funflow} includes both, quantum as well as
thermal fluctuations. For the present purpose we are interested in the
thermal fluctuations, for reviews on thermal FRG see e.g.\
\cite{Litim:1998yn,Litim:1998nf,Blaizot:2009iy}. Thermal fluctuations are encoded in
the difference of the flows at finite and at vanishing temperature,
see e.g.\ \cite{Litim:2006ag},
\begin{equation}\label{eq:thermalflucs} 
  \partial_t \Delta\Gamma_{T,k}[\phi]=\left.\012 \Tr\, G[\phi]
    \cdot \partial_t R\right|_{T}- 
  \left.\012 \Tr\, G[\phi]\cdot \partial_t R\right|_{T=0}\,,
\end{equation}
where 
\begin{eqnarray}\label{eq:DeltaG}
  \Delta\Gamma_{T,k}[\phi]=\left.\Gamma_{k}[\phi]\right|_{T}
  -\left.\Gamma_{k}[\phi]\right|_{T=0}
\end{eqnarray} 
accounts for the difference between the effective action at vanishing
and at finite temperature. Due to the thermal exponential suppression
the flow \eq{eq:thermalflucs} should have locality properties
with respect to the scale $k=T$. As discussed in the previous section, locality
is important for the quantitative reliability of a given
approximation. In the present work we implement this idea as follows:
We use the vacuum physics at vanishing temperature as input for the
flow. A given set of correlation functions $\Gamma^{(n)}_{k=0}$ at
$T=0$ and $k=0$ can be integrated with the flow \eq{eq:funflow} at
vanishing temperature in a given approximation up to a large momentum
scale $k=\Lambda$,
\begin{equation}\label{eq:vaccorr}
  \left.\Gamma^{(n)}_{k=0}(p_1,...,p_n)\right|_{T=0}\stackrel{\rm flow}{\longrightarrow}
 \left. \Gamma^{(n)}_{k=\Lambda}(p_1,...,p_n)\right|_{T=0}\,. 
\end{equation}
In the approximation at hand, \eq{eq:vaccorr} defines the initial
conditions $\left.\Gamma^{(n)}_{k=\Lambda}\right|_{T=0}$, which give the correct
vacuum correlation functions if integrated to $k=0$.  The
ultraviolet scale $\Lambda$ is chosen such that all thermal
fluctuations are suppressed given the maximal temperature to be
considered, $T_{\rm max}$. This implies 
\begin{equation}\label{eq:UVscale}
\0{T_{\rm max}}{\Lambda}\ll 1\,.
\end{equation}  
Then, switching on $T\leq T_{\rm max}$ does not change the initial
conditions at leading order, i.e.
\begin{equation}\label{eq:initialcond}
  \0{1}{\Lambda^{d_n}}\Delta\Gamma^{(n)}_{T,\Lambda}(p_1,...,p_n)=
  0+O\left(\0{T}{\Lambda}\right)\,,
\end{equation} 
where $d_n$ is the canonical dimension of $\Gamma^{(n)}$, and all
momenta are of order $\Lambda$ or bigger, $p_i^2 \gtrsim \Lambda^2$,
see e.g.\ \cite{Blaizot:2010ut} for scalar theories. This suggests
that we can flow $\Delta\Gamma_{T,k}$ from the trivial initial
condition \eq{eq:initialcond} to vanishing cut-off $k=0$. Note also
that within such an approach to thermal fluctuations, it is only the
difference $\partial_t \Delta\Gamma_{T,k}$ which is sensitive to the
approximation at hand. We illustrate the procedure outlined above in a
heuristic plot, see Fig.~\ref{fig:thermalflow}.

In this context it is also worth discussing the rapidity of
$\Delta\Gamma^{(n)}_{T,\Lambda}\to 0$ for large cut-off scales
$k=\Lambda$. This is linked to the question of locality raised in the
previous Section~\ref{sec:local}. In Fig.~\ref{fig:thermalflow} it is
indicated that $\Delta\Gamma^{(n)}_{T,k}$ start to significantly
deviate from zero at the thermal scale $k\approx 2 \pi T$. Above and
in the previous section we have argued that we only have a polynomial
suppression for the flow $ \Delta\dot\Gamma^{(n)}_{T,k}$. Note also,
that the longer $\Delta\Gamma^{(n)}_{T,k}$ survives at large scales
$k\to\Lambda$ the more sensitive it will be to the approximation at
hand. Eventually, the polynomial contributions to the flow integrate up
to the standard thermal exponential suppression, that relates to the
thermal distribution functions as discussed in
Section~\ref{sec:local}.

We conclude that we have a polynomial decay in the flow, which indeed
plays a r$\hat{\rm o}$le for computing quantitatively reliable
thermodynamic quantities. Additionally, we can utilise the ideas
about locality to preserve as much as possible of the exponential
decay, that stabilises any approximation scheme. We turn to the
localised version of the flow derived in the previous
Section~\ref{sec:local}. As the flow of $\hat\Gamma_{k}^{(2)}$
vanishes identically for momenta larger than $\lambda k$ for all
temperatures $T$ we arrive at
\begin{equation}\label{eq:initialcondexp}
\Delta\hat\Gamma^{(2)}_{T,\Lambda}(p)=0+O(e^{-\Lambda/T})\,, 
\end{equation} 
where we have neglected the backreaction of the polynomial decay of
the thermal corrections of higher correlation functions. The two-point
function is the most important quantity if it comes to the computation
of thermodynamic observables. It is exactly here where the locality-preserving
flow pays off in quantitative reliability.

Finally, we are interested in the correlation functions $\Gamma^{(n)}$
at $k=0$, which have to be computed from the $\hat\Gamma^{(n)}$ via the
rescaling with powers of $\hat Z_{k=0}$. At vanishing temperature we
have used the natural normalisation $\hat Z_{k=0}=1$, see
\eq{eq:hatZint} and \eq{eq:hatZ0}, and the two sets of correlation
functions agreed at $k=0$. At finite temperature we initiate the flow
at $k=\Lambda$ with $T/\Lambda$. Here we have \eq{eq:initialcondexp}
for the difference of the localised correlation functions, and we use
the natural definition
\begin{eqnarray}\label{eq:initialhatZT}
\hat{Z}_{\Lambda}(p;T):= \hat{Z}_{\Lambda}(p;T=0)\,. 
\end{eqnarray}
\Eq{eq:initialhatZT} implies that \eq{eq:initialcondexp} also applies
to $\Delta\Gamma_{T,\Lambda}^{(2)}$: at the UV scale we have
exponentially suppressed thermal fluctuations for the two-point
function. As we expect a polynomial suppression due to the arguments
in Section~\ref{sec:local}, the use of the computational simple
initial condition amounts to a temperature-dependent renormalisation
of $\Gamma_k^{(n)}(T)$ at non-vanishing cut-off scale $k$. For the
same reason, \eq{eq:initialhatZT} also leads to $Z_{k=0}(p;T)\neq 1$,
and we have to compute it from the flow of $\hat \eta_k$. We have
\begin{eqnarray}\nonumber 
\hspace{-.3cm}\hat{Z}_{k}(p;T) &=& \hat{Z}_{\Lambda}(p;0)\, \exp 
\left\{ \int_{\Lambda}^{k}dt'\,\hat{\eta}_{k} (p;T)\right\}\\[1ex]
&=& e^{\int_0^k dt'\, \hat{\eta}_{k} (p;0)}\, e^{ \int_{\Lambda}^{k}dt'\,\left(\hat{\eta}_{k} (p;T)-\hat{\eta}_{k} (p;0)\right)}\,,
\label{eq:hatZT} \end{eqnarray}
where we have used \eq{eq:hatZint} in the second line of \eq{eq:hatZT}. 
The relation 
\eq{eq:hatZT} entails that the rescaling factor $Z_{k=0}(p;T)$ contains the thermal part of the 
momentum transfer.

\section{Approximation}\label{sec:approximation}

In this section we discuss the approximation scheme used to capture
the thermal fluctuations and thus allows for a quantitative
computation of the temperature-dependence of the propagators.

In general, the structure of the flow of the effective action, see
Figs.~\ref{fig:funflow},\ref{fig:derG}, entails that flows of
$n$-point functions depend explicitly on $(n\!+\!2)$-point
functions. The relevant example for the present work is the flow of
the (inverse) Yang-Mills propagators $\Gamma_{k}^{(2)}$, displayed in
Fig.~\ref{fig:YM_props}. The flows depends on $\Gamma_{k}^{(n)}$
with $n \leq 4$.  In other words, the functional flow in
eq. (\ref{eq:funflow}), if broken up in the flows of $n$-point
functions, constitutes an infinite hierarchy of coupled
integro-differential equations. For computational purposes the system
must be closed: the set of potentially contributing operators must be
rendered finite, such that the relevant physics is kept in the
approximation. Moreover, the approximation should be subject to
self-consistency checks that give access to the systematic error.

First, let us describe the approximation put forward here for the flow
of the propagators in Fig.~\ref{fig:YM_props}, before we put down the
explicit parameterisation/approximation in terms of $\Gamma_{k}^{(n)}$
with $n\leq 4$: we keep the full propagators and work with
self-consistent approximations to the vertices which respect the
renormalisation group properties of the vertices. We also use the flow
equation for the ghost-gluon vertex, evaluated at the symmetric point
with $p_i^2=k^2$ , $i=1,2,3$, for the momenta of ghost, anti-ghost
and gluon, respectively. We have checked the reliability of the RG-consistent ansatz
for the three-gluon vertex with its flow at the symmetric point
$p^2=k^2$. The reliability of the RG-consistent four-gluon vertex is
checked by its DSE computed in \cite{Kellermann:2008iw}.

\subsection{Two-point functions and their flows}
Now we proceed with the parameterisation of our approximation.  We
concentrate on the localised two-point functions $\hat\Gamma^{(2)}$,
that of the cut-off independent fields are then obtained with
\eq{eq:hatunhat}, \eq{eq:hatZint} and \eq{eq:hatZT}. At vanishing
temperature the propagators are described by one wave-function
renormalisation each, $Z_{A,k}(p)$ and $Z_{c,k}(p)$. At non-vanishing
temperature we have to take into account chromoelectric and
chromomagnetic modes, the respective projection operators $P^L$ and
$P^T$,
\begin{eqnarray}
   P^{T}_{\mu \nu}(p_0, \vec{p}) &=& \left(1-\delta_{\mu 0} \right)
\left(1-\delta_{\nu 0} \right) \left( \delta_{\mu \nu}- p_{\mu}p_{\nu}/\vec{p}^{\ 2} \right),
  \nn\\
  P^{L}_{\mu \nu}(p_0, \vec{p}) &=& \Pi^T_{\mu \nu}(p)-P^{T}_{\mu \nu}(p_0, \vec{p})\,, 
  \label{eq:projections} \end{eqnarray}
where $\Pi^T_{\mu\nu}$ is the four-dimensional transversal projection operator, see \eq{eq:projections0}. 
Thus we parameterise the gluon with two
wave-function renormalisations $Z_{L/T}$. The ghost has only a scalar
structure at vanishing and finite temperature. The parameterisation of
the gluons and ghost is given by
\begin{eqnarray} 
  \hat\Gamma_{A,L}^{(2)}(p_0,\vec{p}) &=&
  Z_{L}(p_0, \vec{p})\, p^2\,P^{L}(p_0,
  \vec{p})\,, \nn\\
  \hat \Gamma_{A,T}^{(2)}(p_0,\vec{p}) &=& Z_{T}(p_0, \vec{p})\, p^2\,P^{T}(p_0,
  \vec{p})\,,
  \nn \\
  \hat \Gamma_{c}^{(2)}(p_0,\vec{p})&= & Z_c(p_0, \vec{p})\,p^2\,, 
 \label{eq:parahatG}\end{eqnarray}
where the identity in colour space is suppressed and the $Z's$ are
functions of $p_0$ and $\vec p$ separately.  The parameterisation of
$\Gamma^{(2)}$ follows from that of $\hat\Gamma^{(2)}$ in
\eq{eq:parahatG}, and is read off the definition of
$\phi(\varphi)$ in \eq{eq:rescaling} and \eq{eq:paraGform},
\begin{equation} \label{eq:paraG}
\Gamma_{A,L/T}^{(2)}\simeq \hat Z_{L/T}(p)\, Z_{L/T}(p)\,p^2\,,\quad  
\Gamma_{c}^{(2)}\simeq  \hat Z_{c}(p)\, Z_{c}(p)\,p^2\,.
\end{equation} 
The flow equations for the two-point functions have been given
diagrammatically in Fig.~\ref{fig:YM_props}. Their right hand sides
depend on the two-point functions, as well as three- and four-point
functions. In particular, we have tadpole diagrams which depend on the
ghost-ghost and ghost-gluon scattering vertices $\Gamma^{(4)}_{\bar c
  c\bar c c}$ and $\Gamma^{(4)}_{\bar c A^2 c}$, respectively. These
vertices vanish classically, and in a first approximation one is
tempted to drop the related diagrams. 

\begin{figure}[t]
\includegraphics[width=\columnwidth]{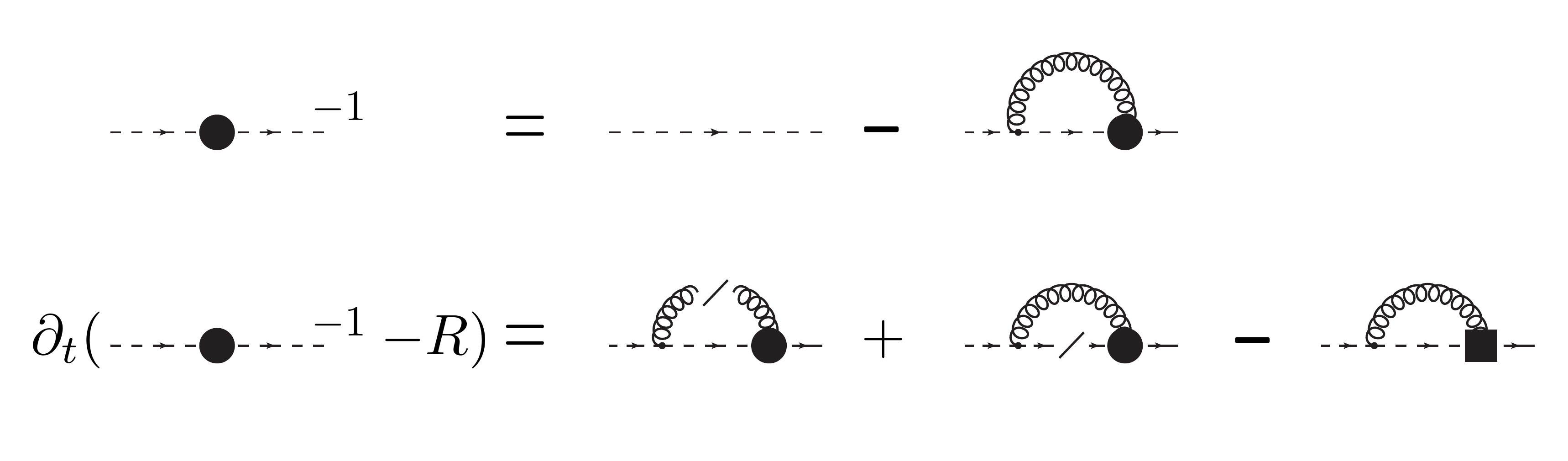}
\caption{Flow equation of the ghost propagator from the total
  $t$-derivative of the renormalised ghost-DSE. The cutted line stands
  for the scale derivative acting on the propagator of the
  corresponding field, see \eq{eq:singlescaleG}. The square denotes
  the scale derivative acting on the full vertex, the vertex without
  filled circle denotes the classical vertex.}
\label{fig:deriv_ghDSE}
\end{figure}

However, they  can be considered in a rather simple way, which we use
for the flow of the (inverse) ghost propagator: we insert the
DSE-relations for $\Gamma^{(4)}_{\bar c c\bar c c}$ and
$\Gamma^{(4)}_{\bar c A^2 c}$ in the related diagrams. This provides a
DSE-resummation of the vertices in a given approximation to the
flow. After some straightforward but tedious algebraic computations, it
can be shown that this turns the flow equation for the ghost in
Fig.~\ref{fig:YM_props} into the total $t$-derivative of the
DSE-equation for the ghost, see Fig.~\ref{fig:deriv_ghDSE}.  This is
nothing but the statement that a flow equation for a correlation
function can be seen as the differential form of the corresponding DSE
for the ghost two-point function in the presence of the regulator
term; both describe the same correlation function. The ghost-DSE and its
$t$-derivative is illustrated in Fig.~\ref{fig:deriv_ghDSE}. The
derivative acting on the dressed propagator gives
\begin{eqnarray}\label{eq:singlescaleG}
\partial_t G\left[ \phi\right] = - G\left[\phi \right]\cdot  
\partial_t \left( \Gamma^{(2)}\left[\phi \right] + R_{\phi}\right)\cdot G\left[\phi \right]\,. 
\end{eqnarray}
The derivative of the bare propagator in the DSE vanishes.
The DSE-flow is finite by construction, as it can be derived from the
manifestly finite ghost flow in Fig.~\ref{fig:YM_props} by inserting
the DSEs for the ghost-ghost and ghost-gluon scattering
vertices. These DSEs are also manifestly finite and require no
renormalisation.  To see this explicitly in Fig.~\ref{fig:YM_props} we
discuss the single terms. First, we note that the total $t$-derivative of the 
propagators, $\partial_t G$, shown in \eq{eq:singlescaleG} act as
UV-regularisation of the loops. They decay at least with $G^2$ as
$\partial_t \Gamma_k^{(2)}$ tends towards a constant for large momenta. The
total $t$-derivative $\partial_t \hat G$ decays even more rapidly
with $\partial_t\hat R$ for large momenta due to \eq{eq:locality}. This
reflects again the locality implemented by \eq{eq:locality} and its
practical importance. The last diagram is proportional to the flow of
the ghost-gluon vertex. It will be discussed in detail in the 
Section~\ref{subsec:ghost-gluon} and is displayed schematically in
Fig.~\ref{fig:cAcflow}. Here it suffices to say that the vertex itself
is protected from renormalisation and its flow decays rapidly for
large momenta.

Both flow equations given in Fig.~\ref{fig:YM_props} and
Fig.~\ref{fig:deriv_ghDSE} for the ghost are exact and are related via
the above resummation procedure. In the present approximation scheme
the DSE is more favorable as it only depends on the ghost-gluon vertex,
which we shall resolve via its flow. In turn, the four-point
functions do not appear explicitly in the total derivative of the DSE,
and the related contributions are absorbed in the diagrams with cutted
propagators. Notably, the ghost-ghost scattering vertex
$\Gamma^{(4)}_{\bar c c\bar c c}$ disappeared completely from the set
of flow equations of ghost and gluon propagators, whereas the
ghost-gluon scattering vertex $\Gamma^{(4)}_{\bar c A^2 c}$ is still
present in the flow of the gluon two-point function.

Note that a similar procedure in the flow of the gluon two-point
function leads to two-loop diagrams, which are commonly dropped in
DSE-computations. This is one of the reasons why we refrain from
resumming the related diagrams.

\subsection{Vertices and their flows}
For the vertices we first introduce a convenient parameterisation that
naturally captures the renormalisation group behaviour. To that end, we
utilise the wave function renormalisation of the gluon and the ghost and write 
\begin{eqnarray} 
  \hat \Gamma^{(n)}(p_1,...,p_{n})= \prod_{i=1}^{n} 
\bar Z^{1/2}_{\phi_i}(p_i)\, {\cal T}(p_1,...,p_{n})\,. 
\label{eq:Gmnsol}\end{eqnarray} 
The $\bar Z$-factors are chosen to be proportional to $Z$, and hence
carry the RG-scaling of the vertex as well as the momentum dependence
of the legs: they carry potential kinematical singularities, see e.g.\
\cite{Fischer:2009tn}. We choose
\begin{eqnarray}\nonumber
\bar Z_{L/T}(p)&=&\0{Z_{L/T}(p)\,p^2-\left[Z_{L/T}(q)q^2\right]_{q=0}}{p^2}\,,\\[1ex]
\bar Z_{c}(p)&=&Z_{c}(p)\,. 
 \label{eq:barZ}\end{eqnarray} 
However, the $\bar Z_{L/T}(p)$ are frozen for $p\leq p_{\rm peak}$,
where $p_{\rm peak}$ is the potential turning point of the inverse
propagator $\Gamma_k^{(2)}$ in the infrared defined by $\partial_p
(p^2 Z_{L/T}(p))_{p=p_{\rm peak}}=0$. The turning point $p_{\rm peak}$
depends on $k$ and tends towards zero for $k\to\infty$. Without this
additional constraint $\bar Z_{L/T}$ would turn negative for small $k$,
which reflects positivity violation. At finite temperature the turning
points depend on $T$ and differ for $Z_T$ and $Z_L$, the turning point
of the latter tends towards $zero$ for $T\to\infty$ due to the Debye
screening. We emphasise that this is done simply for convenience in
order to avoid the splitting of a positive vertex dressing into two
negative factors. The $\bar Z$`s take into account the RG-scaling of the
fields, and reflect the gaps present in the gluonic degrees of
freedom.

This leaves us with a renormalisation group invariant tensor $\cal
T$. It is regular up to logarithms and carries the canonical momentum
dimension as well as the tensor and colour structure. For the flow of
the propagators, Fig.~\ref{fig:YM_props}, $\CT$ has to be computed for
the three-gluon and four-gluon vertex, the ghost-gluon vertex, as well
as for the four-ghost and ghost-gluon scattering vertices. The latter
two, which are absent on the classical level, are treated in
terms of exact resummations with the help of DSEs.  Now we utilise the
locality of the flow: it only carries momenta $q^2\lesssim k^2$ and is
peaked at about $p^2 \approx k^2$. Hence, we approximate the vertices
by evaluating them at the symmetric point at $p_i^2=k^2$ and
vanishing temporal components,
\begin{eqnarray}\label{eq:sympoint}
(p_i)_0^2=0\,\qquad {\rm and}\qquad  \vec p_i^{\,2}=k^2\,.
\end{eqnarray} 
Then the $\bar Z$-factors in \eq{eq:barZ} can be evaluated at fixed
momenta $p$ with \eq{eq:sympoint} and we set
\begin{eqnarray}\nonumber
\bar Z_{k,A}&=&\bar Z_{A}(k)\theta(k-k_s)+Z_{A}(k_s)\theta(k_s-k)\,,\\
\bar Z_{k,C}&=&Z_{C}(k)\,, 
\label{eq:barZk}\end{eqnarray} 
where $Z_{k,A}$ is either $Z_{k,L}$ and $Z_{k,T}$, depending on the
projection $P^{T/L}$ defined in \eq{eq:projections} on the respective
leg of $\Gamma_k^{(n)}$. In a slight abuse of notation we have
introduced $Z(k)$: the $Z$-factors in \eq{eq:barZk} are functions of
$p_0^2$ and $\vec p^2$, which are evaluated at \eq{eq:sympoint}. The
freezing scale $k_s\propto \Lambda_{\rm QCD}$ in \eq{eq:barZk} is
chosen such that it is bigger than $p_{\rm peak}$. We point out that
$k_s$ only defines the parameterisation of the ghost-gluon vertex.

\subsection{Ghost-gluon vertex}\label{subsec:ghost-gluon}
\begin{figure}[t]
\includegraphics[width=.2\columnwidth]{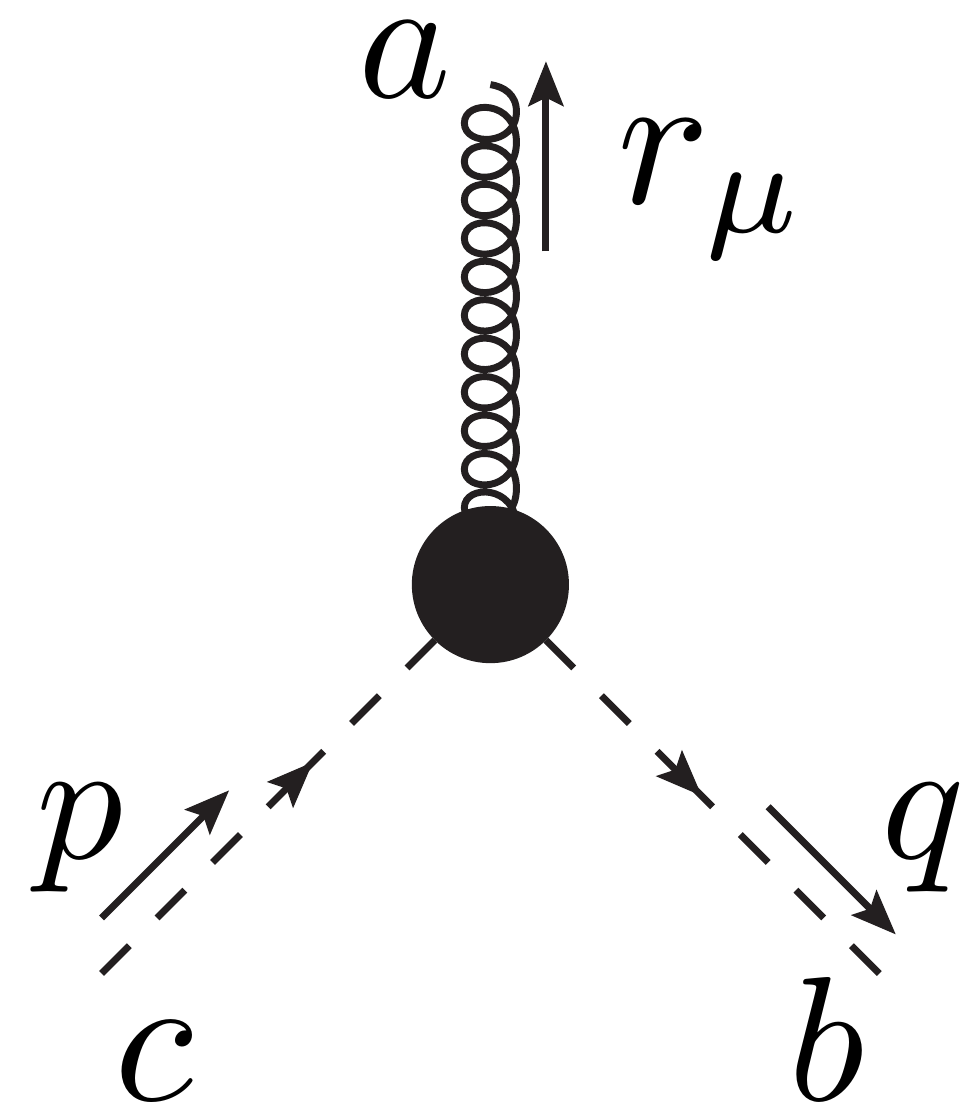}
\caption{Ghost-gluon vertex.}
\label{fig:cAc}
\end{figure}

It is left to determine the dressings $\cal T$ for the primitively
divergent vertices $\Gamma^{(3)}_{A^3}, \Gamma^{(3)}_{A^4}$ and
$\Gamma^{(3)}_{\bar c Ac}$, which have a classical counterpart. We
restrict ourselves to the classical vertex structure. The evaluation
at the symmetric point \eq{eq:sympoint} leaves us with
$k$-dependent dressing functions. For the ghost-gluon vertex we are
led to
\begin{equation} 
  \CT_{\bar cAc,\mu}^{abc}(q,p) =z_{k,\bar c A c}\,\0{1}{g} [S^{(3)}_{\bar cAc}(q,p)
  ]^{abc}_\mu= z_{k,\bar c A c}
 \,i q_\mu f^{abc}\,, 
\label{eq:gglapprox}\end{equation}  
where $g$ is the classical coupling, $S^{(3)}_{\bar cAc}$ is the
classical ghost-gluon vertex derived from \eq{eq:fixedaction}, and
$p$, $q$ are the ghost and anti-ghost momenta respectively, see
Fig.~\ref{fig:cAc}.
The $k$-dependent factor $z_{k,\bar c A c}$ is RG-invariant and defines
a running coupling 
\begin{equation}\label{eq:baralpha}
\bar\alpha_s(k)=\0{z^2_{k,\bar c A c}}{4 \pi}\,,
\end{equation} 
 with running
momentum scale $k$. If expanded in powers of the coupling for large
momenta, $\bar\alpha_s$ has the one and two-loop universal
coefficients of the $\beta$-function of Yang-Mills theory, where we
have used that $\bar Z_{k,A}\to Z_{k,A}$ for large cut-off scales. 

The flow of $z_{\bar cAc}$ is extracted from that of the ghost-gluon vertex.
Here this flow is computed within a DSE-resummation
similar to the derivation made for the ghost propagator.
\begin{figure}[t]
\includegraphics[width=.9\columnwidth]{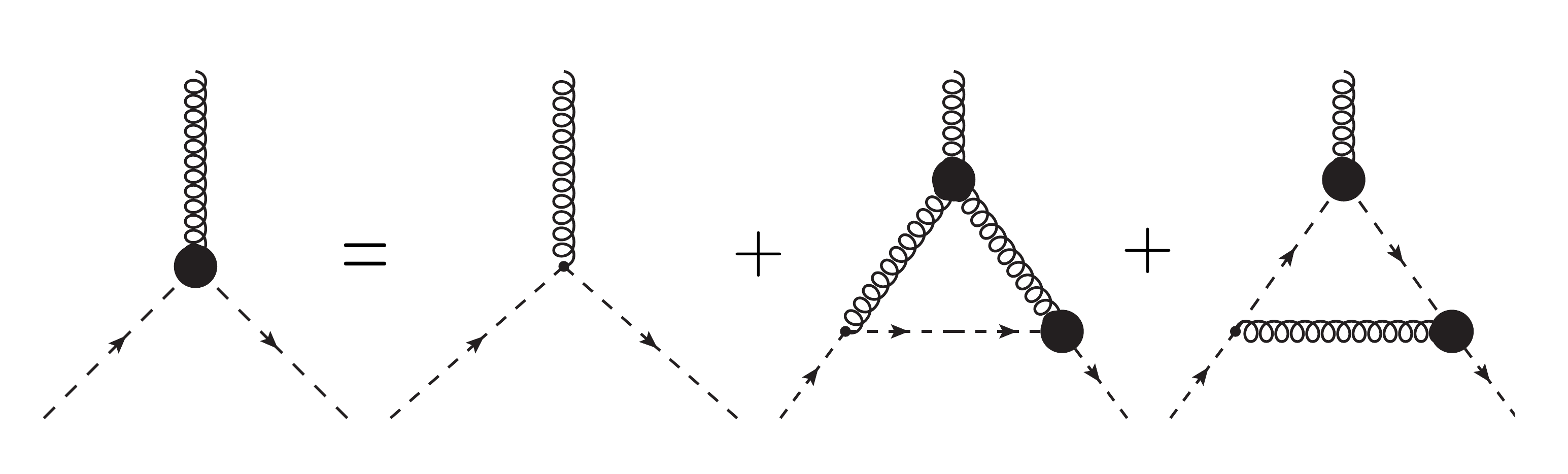}
\caption{DSE for the ghost-gluon vertex. The vertex without filled
  circle denotes the classical vertex.}
\label{fig:cAc_DSE}
\end{figure}
Again, the DSE-resummed flow is finite as it is derived from the
standard flow equation for the ghost-gluon vertex which is finite by
construction. Note however, that already the DSE in
Fig.~\ref{fig:cAc_DSE} is finite without renormalisation procedure due
to the non-renormalisation theorem for the ghost-gluon vertex. Due to
simple kinematical reasons, it is also present in approximation
schemes that respect the kinematical symmetries.

The above arguments allow us to start straightaway with the simple
ghost-gluon DSE, see Fig.~\ref{fig:cAc_DSE}, which only contains one-loop
terms. Similarly to the flow
for the ghost propagator, we turn the DSE in Fig.~\ref{fig:cAc_DSE}
into a flow equation by taking the $t$-derivative of
Fig.~\ref{fig:cAc_DSE}. The running of the vertex allows for a temperature-dependent computation
of the vertex, which will have an important effect in the calculation
of the propagators. 

It remains to project the vertex flow onto that of the renormalisation
group invariant dressing function $z_{k,\bar c Ac}$. The projection on
the classical vertex structure is done by
\begin{equation}\label{eq:GbarcAcps}
 \Gamma_{\bar c Ac}(p_s) = \left(\0{[\hat
      \Gamma_{\bar c Ac}^{(3)}]_\mu^{abc}\, [S^{(3)}_{\bar cAc}]_\mu^{abc}}
    {[S^{(3)}_{\bar cAc}]_\nu^{def} [S^{(3)}_{\bar cAc}]_\nu^{def}}
\right)_{p^2=q^2=(p+q)^2=p_s^2}\,,  
\end{equation} 
with an evaluation at the symmetric point at the momentum scale
$p_s$. For the classical vertex $\Gamma_{\bar c Ac}^{(3)}=S_{\bar c
  Ac}^{(3)}$ derived from \eq{eq:fixedaction}, the dressing is simply
unity, $\Gamma_{k,\bar c Ac}(p_s) =1$. In the present approximation we
evaluate the vertices at the moment scale $k$, and hence we define
\begin{eqnarray}\label{eq:GbarcAc}
\Gamma_{k,\bar c Ac}=\Gamma_{\bar c Ac}(k)\,. 
\end{eqnarray}
Note that the lhs depends on $k$ via the evaluation at $p_s=k$ but
also due to the implicit dependence of the vertex on the cut-off
scale. The full vertex dressing in \eq{eq:GbarcAc} also includes the
dressing of the legs as split off in \eq{eq:Gmnsol}. Hence, the
dressing function $z_{k,\bar c Ac}$ is given by
\begin{equation}\label{eq:zbarcAc}
  z_{k,\bar c Ac} =\0{1}{\bar Z_{k,A}^{1/2} \bar Z_{k,c}} \Gamma_{k,\bar c Ac}\,.
\end{equation}
The flow $\partial_t z_{k,\bar c Ac} $ is determined from 
\eq{eq:zbarcAc} and is directly related to that of the ghost-gluon
vertex. Taking the $t$-derivative of \eq{eq:zbarcAc} leaves us with
\begin{eqnarray}\nonumber 
  \left(\partial_t+\012 \0{\partial_t \bar Z_{k,A}}{\bar Z_{k,A}}+
    \0{\partial_t \bar Z_{k,c}}{\bar Z_{k,c}}\right) z_{k,\bar c Ac}=\partial_t 
\Gamma_{k,\bar c Ac}\, \0{1}{\bar Z_{k,A}^{1/2} \bar Z_{k,c}} \,.
\label{eq:flowzbarcAc}\end{eqnarray}
The scale-derivative of the full dressing $\Gamma_{\bar c Ac}$ is
proportional to the flow of the vertex but also to the derivative with
respect to the momentum at the symmetric point,
\begin{eqnarray}\nonumber 
 \partial_t \Gamma_{k,\bar c Ac} = \left[\partial_t \Gamma_{\bar c Ac}(p_s)+ 
 p_s \partial_{p_s} \Gamma_{\bar c Ac}(p_s)\right]_{p_s=k} \,.
\label{eq:flowGbarcAc}\end{eqnarray}
Upon integration the flow \eq{eq:zbarcAc} gives us
the vertex dressing of the ghost-gluon vertex at a given cut-off scale
$k$,
\begin{equation} z_{k,\bar c Ac} =
z_{k=0,\bar c Ac}+ \int_0^{k}\frac{dk'}{k'} \partial_{t'} z_{k',\bar c Ac}.
\end{equation} 
For the thermal flows in the present work the initial condition
$\left.z_{k=0,\bar c Ac}\right|_{T=0}$ is required. It is here where
we take advantage of the progress made over the last two decades in
our understanding of Landau gauge QCD at vanishing temperature, for a
review see \cite{Fischer:2008uz}. At vanishing temperature a
one-parameter family of solutions with infrared enhanced ghost
propagators and gapped gluon propagators in Landau gauge has been
found. All solutions have a gluon propagator with a mass gap $m_{\rm
  gluon}\propto \Lambda_{\rm QCD}$. They only differ from each other
in the deep infrared for momenta $p^2\ll\Lambda_{\rm QCD}$. There 
the gluon propagator is described by 
\begin{eqnarray} \label{eq:deepIR} 
Z_A(p\ll \Lambda_{\rm QCD}) \propto c(p) \0{m^2_{\rm gluon}}{p^2}\,, 
\end{eqnarray} 
where $c(p)\gtrsim 1$ is a momentum-dependent function which is
bounded from below. For all solutions but one $c(p)$ it is also
bounded from above, these solutions are called decoupling solutions,
as the gluon decouples in that momentum regime. There is one
distinguished member of this family where $c(p)$ diverges with
$p^{2+2\kappa_A}$ with $\kappa_A<-1$, see \cite{Fischer:2008uz}. This
solution is called scaling solution as the infrared propagators and
vertices are uniquely determined by scaling laws up to constant
prefactors, see e.g.\ 
\cite{Zwanziger:2001kw,Lerche:2002ep,Alkofer:2004it,%
Fischer:2006vf,Fischer:2008uz,Fischer:2009tn,Alkofer:2008jy}.

\begin{figure}[t]
\includegraphics[width=.9\columnwidth]{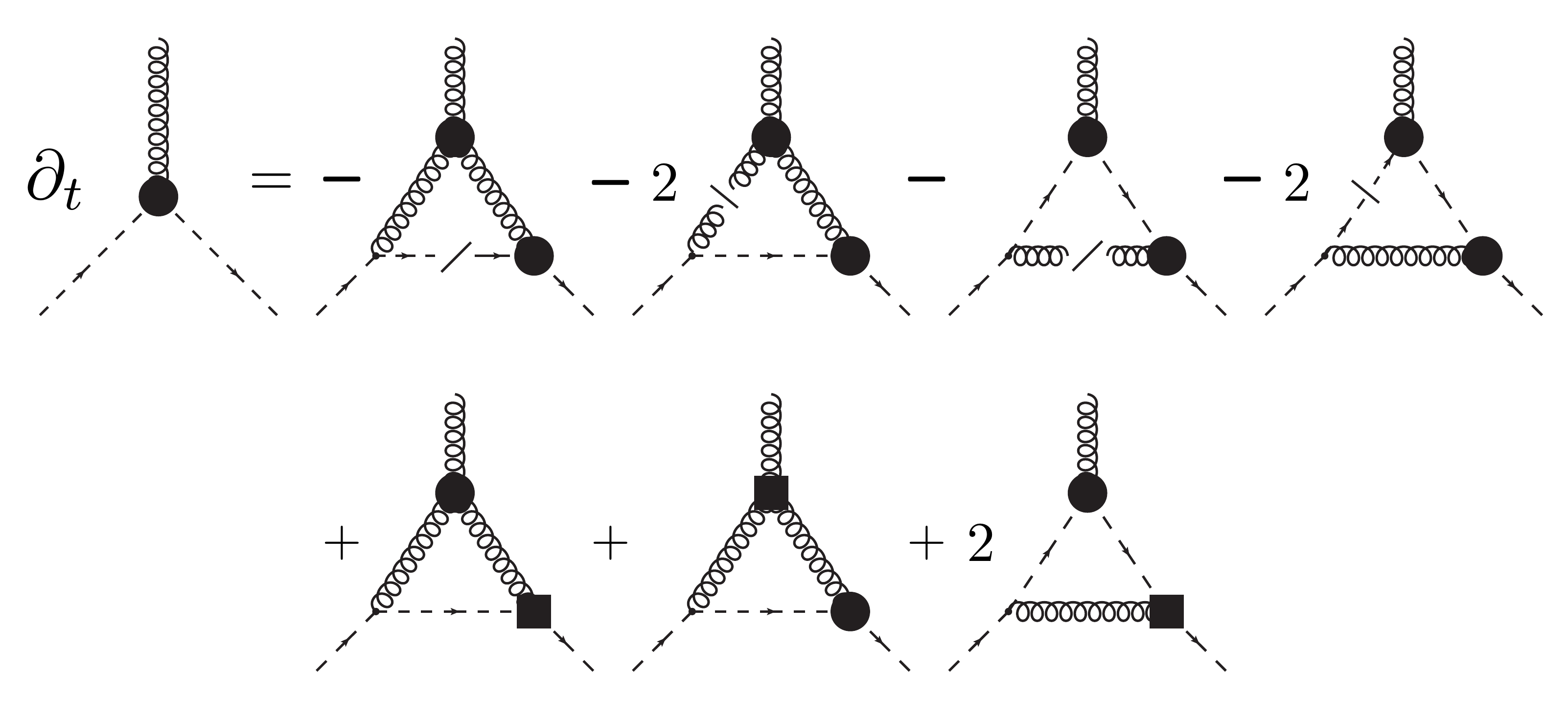}
\caption[]{Flow equation for the ghost-gluon vertex from the total
  $t$-derivative of the ghost-gluon vertex DSE in
  Fig.~\ref{fig:cAc_DSE}. The cutted line stands for the scale
  derivative acting on the propagator of the corresponding field, see
  \eq{eq:singlescaleG}. The square denotes the scale derivative acting
  on the full vertex, the vertex without a filled circle denotes the
  classical vertex. }
\label{fig:cAcflow}
\end{figure}
Most importantly, the flow of the ghost-gluon vertex in
Fig.~\ref{fig:cAcflow} is not sensitive to the infrared behaviour of
the gluon propagator. All diagrams in Fig.~\ref{fig:cAcflow} vanish in
the limit $k\to 0$ with powers of $k^2/m^2_{\rm gap}$, 
\begin{eqnarray} \label{eq:insensitive} 
\lim_{k\to 0}\partial_t z_{k,\bar c Ac} \propto \0{k^2}{m^2_{\rm gap}}\,.
\end{eqnarray}
This also entails that the vertex dressing tends toward a constant in
the infrared. Hence, we infer that the infrared value of the
ghost-gluon vertex is the same for the whole class of solutions up to
subleading terms in RG-transformations. This can be used to
compute $\left.z_{k=0,\bar c Ac}\right|_{T=0}$ for the whole class of
solutions. It has been shown that the scaling solution for constant
ghost-gluon dressings is determined in the FRG up to an RG-constant,
see \cite{Pawlowski:2003hq},
\begin{subequations}\label{eq:IRvalue}
\begin{equation}\label{eq:IRvalue1}
  \left.  z^2_{k=0,\bar c Ac}\right|_{T=0} \left(\0{\bar Z_{A}(p)\bar Z^2_{c}(p)}{
Z_{A,s}(p)Z^2_{c,s}(p)}\right)_{p=0}=
    4 \pi\alpha_{s,\rm IR} \,,
\end{equation} 
with the scaling wave function
renormalisations $Z_{A/C,s}$ for ghost and gluon propagators, respectively.
The coupling $\alpha_{s,\rm IR}$ is analytically known, see \cite{Lerche:2002ep},
\begin{eqnarray}\label{eq:IRvalue2} 
  \alpha_{s} = -\frac{4 \pi}{N_c}\frac{2}{3} \frac{\Gamma(-2\kappa)
    \Gamma(\kappa-1)\Gamma(\kappa+3)}{\left(\Gamma(-\kappa)\right)^2 
    \Gamma(2\kappa-1)} \overset{N_c=3}{\approx} 2.97\,,
\end{eqnarray}  
\end{subequations} 
where $\kappa \approx 0.595$.  

For momenta $p\gg \Lambda_{\rm QCD}$ the $Z_{A/C,s}(p)$ tend towards
the decoupling solutions up to RG-scalings. Demanding equivalence for
large momenta fixes the relative ultraviolet renormalisation
condition. In summary, this allows us to fix $z^2_{k=0,\bar c Ac}$ at
vanishing temperature for a given set of scaling or decoupling
propagators in terms of the UV renormalisation condition.

In the light of the ongoing debate about the infrared behaviour of
Landau gauge propagators in the vacuum ($T=0$) it is important to
stress the following: first, the flow of the vertex function is not
sensitive to the differences of the momentum behaviour of the
propagators in the deep infrared, $p\ll \Lambda_{\rm QCD}$, as it is
switched off for $k\to 0$ below $\Lambda_{\rm QCD}$. Second, the above
argument leading to \eq{eq:IRvalue} only relies on the technical
possibility of finding initial ultraviolet conditions for the flow in
the given approximation which flow into the scaling solution in the
infrared. This is trivially possible, see
\cite{Pawlowski:2003hq,Fischer:2004uk,Fischer:2008uz}. Then, the
analytical values of the scaling solution fixes \eq{eq:IRvalue} for
both, scaling and decoupling solutions. This does not resolve the
infrared problem in Landau gauge Yang-Mills theory closely related to
the picture of confinement, as well as to the resolution of the Gribov
problem in this gauge.

At non-vanishing temperature the coupling is suppressed below 
the temperature scale $k\sim T$, see e.g. \cite{Braun:2005uj}.
Furthermore, we have to distinguish between
transversal and longitudinal gluon legs. If all Matsubara frequencies
vanish, $p_0=q_0=0$, the longitudinal vertex vanishes and the
distinction is only relevant for non-vanishing Matsubara
frequencies. There, however, all (hatted) quantities quickly tend
towards their $T=0$ counter parts. Therefore, we approximate the
longitudinal vertex dressing at finite temperature by the transversal
one, 
\begin{eqnarray}\label{eq:long+trans}
z_{k,\bar c Ac}^L=z_{k,\bar c Ac}^T\,.
\end{eqnarray}
In summary, we have set up a relatively simple flow for the ghost-gluon
vertex which already covers the quantitative features of the full
vertex flow. We have also checked our approximation by computing the
full vertex flow on the basis of the results in the present
approximations. This also provides us with an estimate of the systematic error 
in the present approximation. A full analysis of this will be published elsewhere.

\subsection{Gluonic vertices}\label{subsec:gluonicvertices}

It remains to determine the purely gluonic vertices. They are
described within a parameterisation similar to \eq{eq:gglapprox}. We
have schematically
\begin{eqnarray} 
\CT_{A^3} = z_{k,A^3}\,\0{1}{g} S^{(3)}_{A^3} \,, 
\qquad \CT_{A^4} = z_{k,A^4}\,\0{1}{g^2} S^{(4)}_{A^4} \,, 
\end{eqnarray}  
where the vertex dressings $z_{k,A^3}$ and $z_{k,A^4}$ relate directly
to the ghost-gluon dressing $z^2_{k,\bar c Ac}=4 \pi\bar \alpha_s(k)$,
\eq{eq:baralpha}, for large cut-off scales $k\gg \Lambda_{\rm QCD}$ or
large momenta due to two loop universality. 

\begin{figure}[t]
\includegraphics[width=.7\columnwidth]{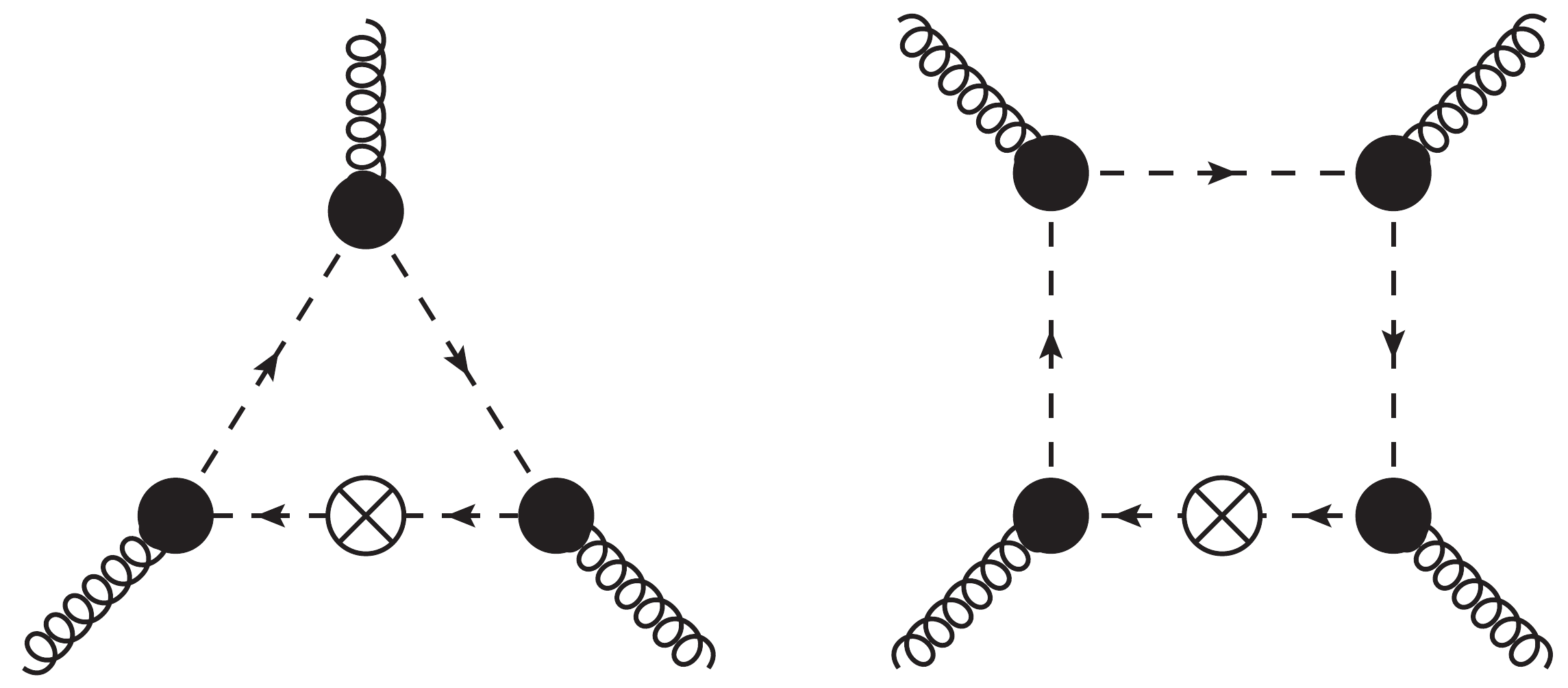}
\caption{Infrared dominating diagrams in the flow equations for the
  three-gluon vertex and the four-gluon vertex.}
\label{fig:ghostloops}
\end{figure}

Indeed, this reasoning has been validated in \cite{Kellermann:2008iw}
with DSE-equations for the four-point coupling. Hence, we have
$z_{k,A^4}\simeq z^2_{k,\bar cAc}$ for most of the momentum
regime with a potential deviation in the deep infrared for 
$k^2/\Lambda_{\rm QCD}^2\ll 1$. We parameterise accordingly
\begin{eqnarray}\label{eq:f}
z_{k,A^3}= z_3\,z_{k,\bar cAc}\,,\quad 
\alpha_{s,A^4}= z_4\, z^2_{k,\bar cAc}\,, 
\end{eqnarray} 
where the $z_i, i=3,4$, are functions which are expected to approach
unity for $k^2\gtrsim \Lambda^2_{\rm QCD}$, that is
\begin{eqnarray}\label{eq:f1}
z_{i,k\gg \Lambda_{\rm QCD}} \to  1\,, 
\end{eqnarray} 
for $i=3,4$. Their infrared behaviour is determined by the only
diagram in the flow that does not depend on the gapped gluon
propagator, see Fig.~\ref{fig:ghostloops}.
The effect of these diagrams on the vertices is determined
by two competing effects:

First, these diagrams are suppressed due to their colour structures
and the related vertex dressings are suppressed relative to that of
the ghost-gluon vertex. In the case of the four-gluon vertex this
combinatorial suppression is of order $10^2-10^3$, which is nicely seen
in the solution in \cite{Kellermann:2008iw}. For the three-gluon
vertex the combinatorial suppression factor also turns out to be of
order $10^2-10^3$, subject to the chosen momenta, details will be
published elsewhere. This factor has been determined within a
two-dimensional lattice study as $0.017$, \cite{Maas:2007uv}. 

Second, these diagrams grow strong in comparison to the respective
diagram in the flow of the ghost-gluon vertex, which involves one
gapped gluonic propagator. In the decoupling case the flow of the
gluonic vertices for $k\to 0$ is proportional to $k^0$ up to
logarithms. This leads to an effective suppression of diagrams with
gluonic vertices as such diagrams also involve gapped gluon
propagators, and hence a suppression with $k^2/m^2_{\rm gap}$. In the
scaling case the diagrams with gluonic vertices decouple with powers
of the scaling, which can be seen best in the DSE-hierarchy.

In summary, the above analysis entails that we can safely drop the
related diagrams for $k\leq \Lambda_{\rm QCD}$, and above this scale
the vertices have a dressing similar to that of the ghost-gluon
vertex. In the vertices we have factorised the multiplicative dressing
functions $\bar Z^{1/2}$ for the respective legs that are at our
disposal. In turn, the relative thermal suppression factor is safely
encoded in the ratio of the full ghost-gluon vertex dressings at
vanishing and finite temperature, 
\begin{eqnarray}\label{eq:ratio}
\0{\left.\Gamma_{k,\bar c Ac}\right|_{T}}{\left.\Gamma_{k,\bar c Ac}\right|_{T=0}}\,.
\end{eqnarray}
Since the temperature dependent
dressing factor $z_{\bar c Ac}$ has already been included in the
definition \eq{eq:f}, we deal with the reduced ratio of the wave
function renormalisations, 
\begin{equation}\label{eq:rbarcAc}
r_{\bar c A c}(k,T)  = \frac{\left.\bar Z_{k,A}^{1/2}\,\bar Z_{k,C}\right|_{T}}
{\left.\bar Z_{k,A}^{1/2}\,\bar Z_{k,C}\right|_{T=0}}\,,
\end{equation} 
leading to the final approximation of the gluonic vertex dressings by
\begin{equation}\label{eq:finaldressing}
  z_i(k,T) = r_{\bar c A c}(k,T)^{i-2}\,, \quad  z^{\rm min}_i(k,T) 
=\0{ \bar Z_{k,A} }{  Z_{k,A} }  z_i(k,T)  \,.
\end{equation}  
The results we show are achieved within the second set of vertex
dressings, where we additionally switch-off the gluonic vertices
linearly below the temperature scale, which implements the additional
thermal suppression in the infrared. This betters the numerical
convergence. In Section~\ref{sec:props+vertices} we
test the sensitivity of the results to this switch-off.

At non-vanishing temperature the structure functions $z_{A^n}$ also
have to carry the difference between the coupling to longitudinal and
transversal gluons, which is relevant for the second choice in
\eq{eq:finaldressing}. The ratio $\bar Z_A/Z_A$ goes to $\bar
Z_{L/T}/Z_{L/T}$. The validity of the present approximation is then
tested by computing the flow of $z_{A^3}$ on the basis of our
results. Note also that $z_{k,\bar cAc}$ only has transversal parts as we
evaluate the flows at vanishing Matsubara frequency.

\subsection{Regulators} 
For a numerical treatment of finite temperature flow equations,
exponential regulators 
\begin{equation}
r_m(x)= \frac{x^{m-1}}{ e^{x^m}-1}
\label{eq:reg}
\end{equation}
generally provide good numerical stability. The parameter $m$ controls
the sharpness of the regulator, which directly relates to the locality
of the flow. As already pointed out in section \ref{sec:local},
locality is a central issue. The importance of sufficiently local
flows is also seen on the level of the regulator, where it turns out
that the non-locality induced by the exponential regulator
(\ref{eq:reg}) with $m=1$ spoils the stability of the flow. On the
other hand, regulators with a steep descent lead to a slower
convergence of the regularised thermal propagators and vertices
towards the vacuum ones at $T=0$. Indeed, for sharp cut-off functions
or non-analytic ones thermal modifications are present for arbitrarily
large cut-off scales $\Lambda$. This invalidates the use of the $T=0$
initial conditions at the initial UV scale $\Lambda$. Accordingly, we
use $m=2$ in the computation, whose form is shown in
Fig. \ref{fig:R}. The full regulators for the gluon and the ghost are
given by
\begin{eqnarray} 
\hat R^{T/L}{}_{\mn}^{ab}(p)& =& \delta^{ab}
  P^{T/L}_{\mn}\,
  \bar Z_{k,T/L}\, p^2 r_2(p^2/k^2)\,,\nn\\
  \hat R^{ab}(p)& =& \delta^{ab}\, Z_{c}(0)\, p^2 r_2(p^2/k^2)\,, 
\end{eqnarray}
with the projection operators $P^{L/T}$ defined in
\eq{eq:projections}. We have chosen $Z_c(0)$ instead of $Z_c(k)$ for
numerical convenience. The ghost renormalisation function $Z_c(p;T)$
tends towards zero for small momenta and finite temperatures. It is a
balance between ghost propagator, gluon propagator and ghost-gluon
vertex, which prohibits $Z_c$ getting negative. This balance is better
resolved numerically with the prefactor $ Z_{c}(0)$ in the regulator
$R_c$ for the ghost. At vanishing temperature we have $\bar Z_{k,L}=\bar Z_{k,T}$, and the
gluon regulator $R_{k,A}=R^T+R^L$ is proportional to the
four-dimensional transversal projection operator $\Pi$ defined in
\eq{eq:projections}, see also the discussion below \eq{eq:dSk}. 

\section{Computational Details}\label{sec:comp}
\begin{figure}[t]
  \includegraphics[width=\columnwidth]{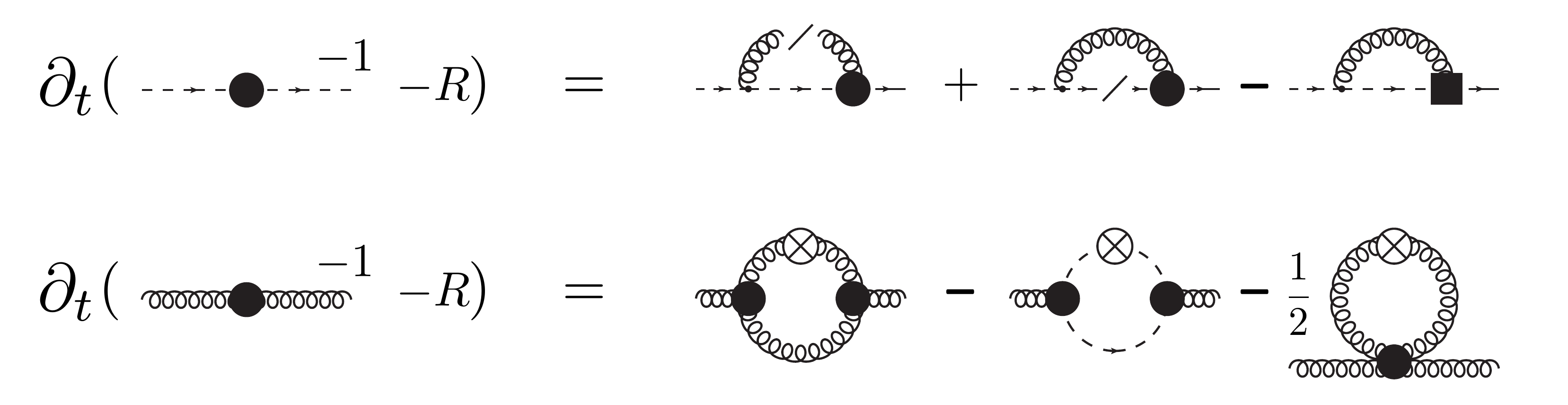}
\includegraphics[width=.9\columnwidth]{figures/DSE_approx_cAc_large_eq}
\caption{Yang-Mills flows for propagators and ghost-gluon vertex in
  the approximation discussed in Section~\ref{sec:approximation}. The flows for
  the ghost propagator and the ghost-gluon vertex are DSE-resummed and
  the ghost-tadpole in the gluon equation is neglected.}
\label{fig:truncation}
\end{figure}

The discussion of the last section leaves us with a coupled set of
partial integro-differential equation depicted in
Fig.~\ref{fig:truncation}. The initial condition is set in the vacuum,
$T=0$, at vanishing cut-off scale, $k=0$. The vertices are determined
by \eq{eq:IRvalue2} and the relations in the
  Sections~\ref{subsec:ghost-gluon},\ref{subsec:gluonicvertices} and
  the input gluon and ghost propagators are taken from
  \cite{Fischer:2008uz}, from which we choose a decoupling solution.

In order to solve the flow equation up to a UV scale $\Lambda$ we
adopt an iteration procedure, which is very stable for the direction
$k\ra\Lambda$. Herein the iteration $(i+1)$ only depends on the given
solution $(i)$, 
\begin{eqnarray}\label{eq:iteration} &&\Gamma^{(n)}_{k,i+1}=
\Gamma^{(n)}_{k=0,i}+\nn\\
&&\hspace{1.5cm}+ \int_{0}^{k}\frac{dk'}{k'}
\tn{Flow}^{(n)}_{i+1}\left(\Gamma^{(n)}_{k,i},\tn{Flow}^{(n)}_{i}\right)\,.
\end{eqnarray} 
Incrementing the number of iterations until the solution is stable
under further iteration,
i.e. $\Gamma^{(n)}_{k,m+1}=\Gamma^{(n)}_{k,m}$ up to desired accuracy,
gives the solution of the flow equation.  The starting point of the
iteration we take as $\Gamma^{(n)}_{k,(0)} =
\Gamma^{(n)}_{T=0,k=0}$. Then we converge within the iterations
rapidly to the initial condition $\Gamma^{(n)}_{\Lambda,T=0}(p^2)$.

In contrast to the direction $k\ra \Lambda$ in the zero temperature
case, the flow from high to low scales at finite temperature involves
instabilities. This is due to the self-regulating nature of the
equations for the wave-function renormalisation of the ghost in
combination with the ghost-gluon vertex dressing. Their structure is
such that if one of the quantities becomes small it stops the flow of
the other and in the following also its own flow, which is what
happens at finite temperature. However, as soon as one of these
quantities happens to be negative in an intermediate iteration step,
the iteration becomes unstable, i.e. each iteration step brings the
iteration solution further away from the correct solution. The system
is highly sensitive to this, as one has to resolve very small values
numerically. Therefore, we pursue a more direct strategy to solve the
flow, namely an evolution of the flow according to a Runge-Kutta
solver \beqa &&\Gamma^{(n)}_{k_{i-1}} = \Gamma^{(n)}_{k_{i}}+
\frac{k_{i-1}-k_{i}}{k_{i}} \tn{Flow}^{(n)}_{k_{i}}, \eeqa from the
starting condition $k_N = \Lambda\gg T$ to $k_0=0$.  In the evolution
the system reacts on the balancing effect between the ghost propagator
and ghost-gluon vertex immediately, and the purely numerical problem
of possibly negative values of $Z_c$ or $z_{\bar{c}Ac}$ in the
iteration is avoided. Note that the result is stable under iteration
again, as it is the exact solution of the equation.  The sensitivity
to the balancing is still present, but shows up in the form of a small
evolution step size of $\left|k_{i-1}-k_{i}\right| \lesssim
10\tn{MeV}$.

\section{Results for Propagators and Vertices}\label{sec:props+vertices}

In this section we present the results for the ghost and gluon
propagators and the ghost-gluon vertex. The temperature is given in lattice units. As has been argued in
section~\ref{sec:thermalflow}, the results for the ghost and gluon
propagators show the typical thermal scale $ 2 \pi T$. Below this
scale we have significant temperature effects on the momentum
dependence. In turn, above this scale the temperature fluctuations are
suppressed and all propagators tend towards their vacuum counterparts
at vanishing temperature. This also holds true for the ghost-gluon
vertex. This supports the self-consistency and stability of the
thermal flows as discussed in section \ref{sec:thermalflow}.

\begin{figure}[t]
\includegraphics[width=.9\columnwidth]{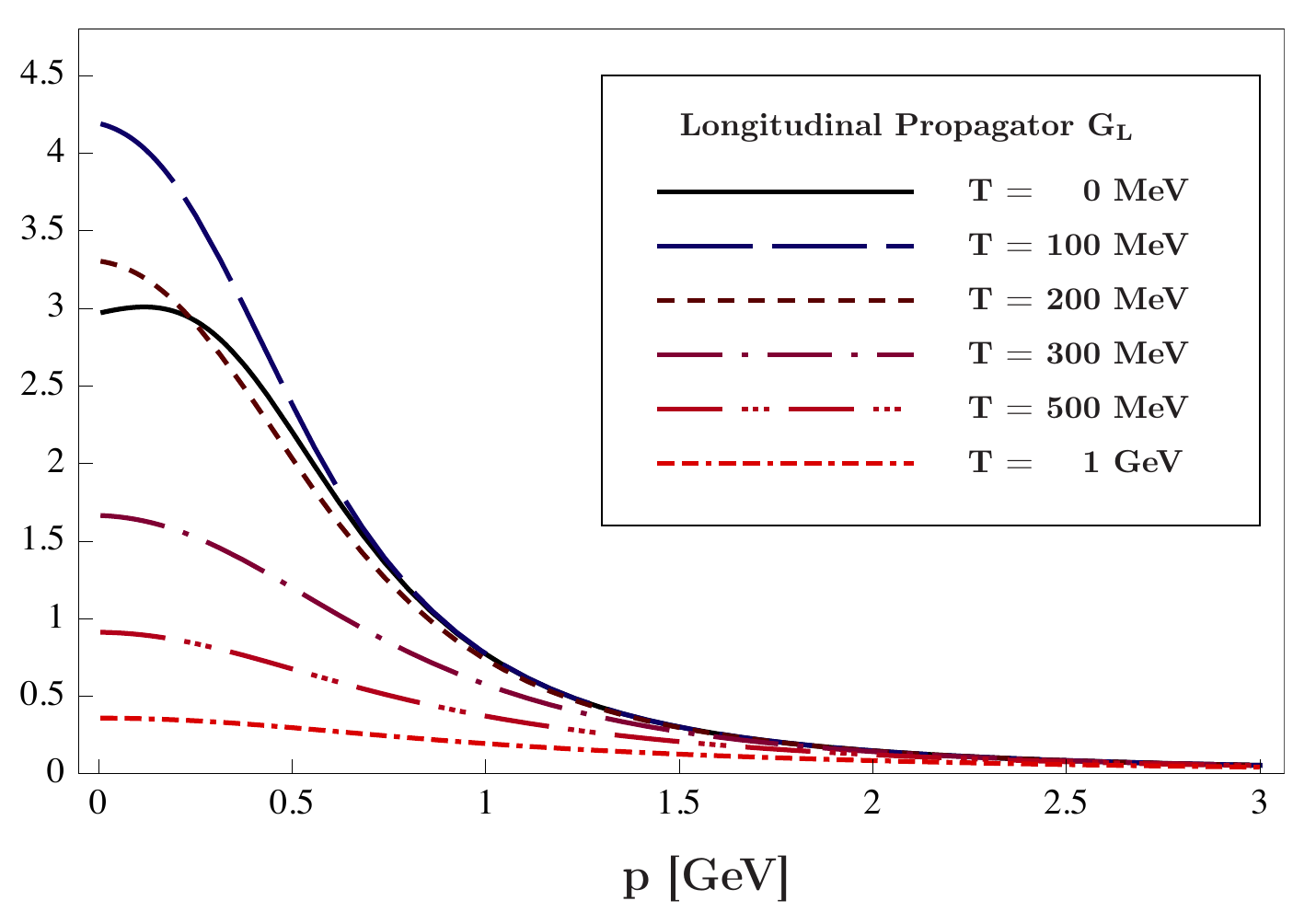}
\caption[]{Longitudinal gluon propagator $G_L$ at different
  temperatures as a function of spatial momentum.}
\label{fig:GL}
\end{figure}

The most significant effect can be seen for the chromoelectric and
chromomagnetic gluon propagators, that is the components of the
propagator longitudinal and transversal to the heat bath.  The zero
mode of the longitudinal gluon propagator at various temperatures is
given in Fig.~\ref{fig:GL} as a function of spatial momentum.
For low temperatures $T\lesssim 150\textnormal{ MeV} $ we see an
enhancement of the longitudinal propagator. Such an enhancement is
also seen on the lattice,
\cite{Maas:2011ez,Aouane:2011fv,Maas:2011se,Cucchieri:2011di,Fischer:2010fx,Cucchieri:2007ta}.
It has been emphasised on the basis of the FRG that specifically the
propagator of the electric mode should show critical behaviour, if
computed in the background fields that solve the non-perturbative
equations of motion \cite{Maas:2011ez}. However, the significance of
the lattice results so far as well as quantitative details are not
settled yet. For higher temperatures the longitudinal propagator is
suppressed relative to the gluon propagator at vanishing temperature.
This is the expected behaviour caused by the Debye screening mass due
to the thermal screening of the chromoelectric gluon. For
asymptotically high temperatures $T\gg T_c$ the chromoelectric gluon
decouples. The onset of this behaviour at about $T\approx
200\textnormal{ MeV}$ is earlier as in the respective lattice
computations
\cite{Maas:2011ez,Aouane:2011fv,Maas:2011se,Cucchieri:2011di,Fischer:2010fx,Cucchieri:2007ta}
where the thermal decoupling takes place for temperatures larger than
the critical temperature.

\begin{figure}[t]
\includegraphics[width=.9\columnwidth]{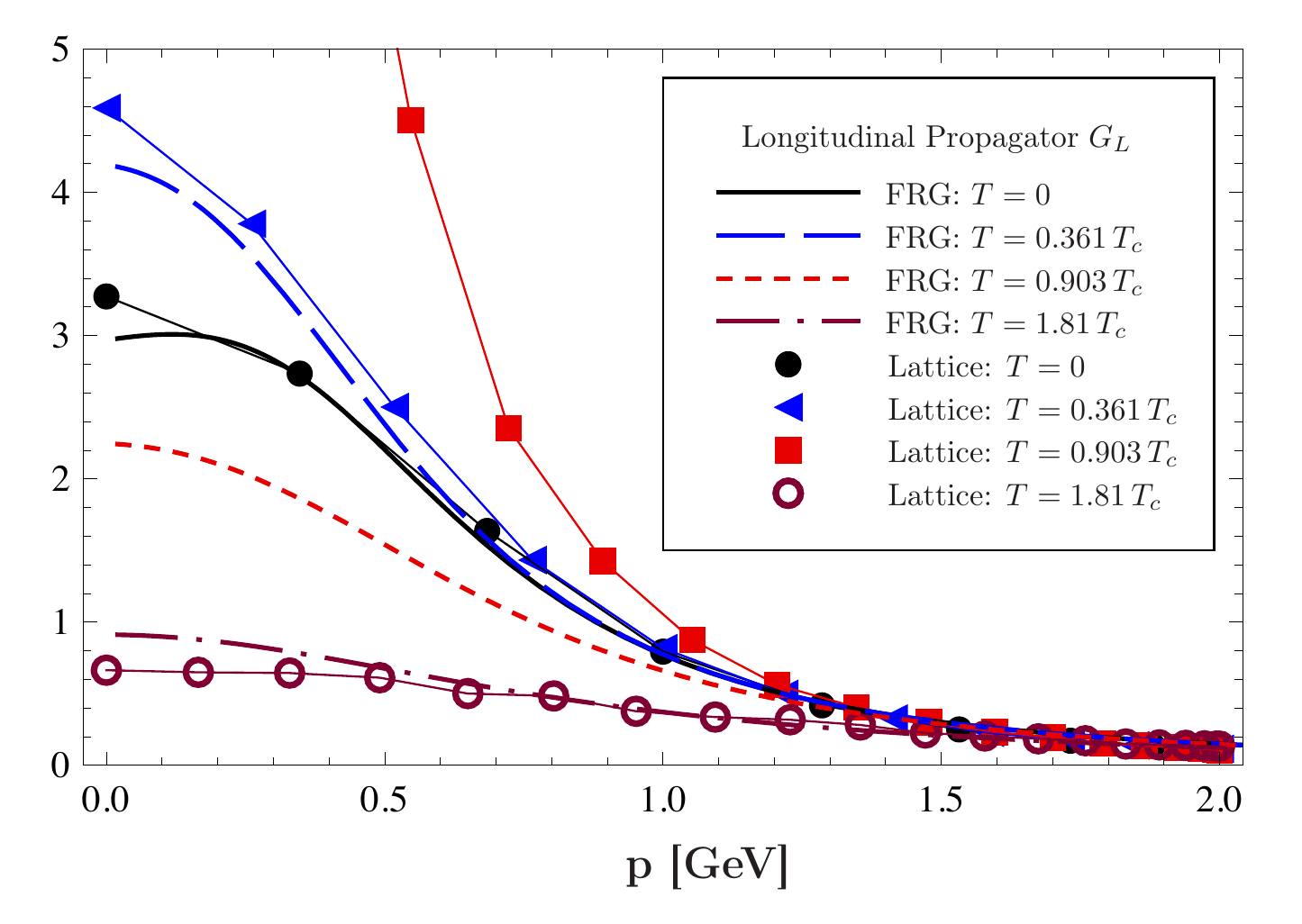}
\caption[]{Longitudinal gluon propagator in comparison with lattice
  results \cite{Fischer:2009gk}. The lattice data has been rescaled
  such that the $T=0$ propagators match at intermediate momenta
  $p\gtrsim 1\tn{ GeV}$.}
\label{fig:FRG_latt_long}
\end{figure}

In order to capture this behaviour we have to extend our present
truncation with a self-consistent inclusion of the Polyakov loop
background as well as a better resolution of the purely gluonic
vertices for momenta and frequencies below $\Lambda_{\rm QCD}$. Both
extensions are under way, and the results will be presented elsewhere.
In turn, for large temperature and momentum scales above $\Lambda_{\rm
  QCD}$ the above lack of quantitative precision at infrared scales is
irrelevant. Here we see quantitative agreement with the lattice
results, see Fig.~\ref{fig:FRG_latt_long}.

\begin{figure}[t]
\includegraphics[width=.9\columnwidth]{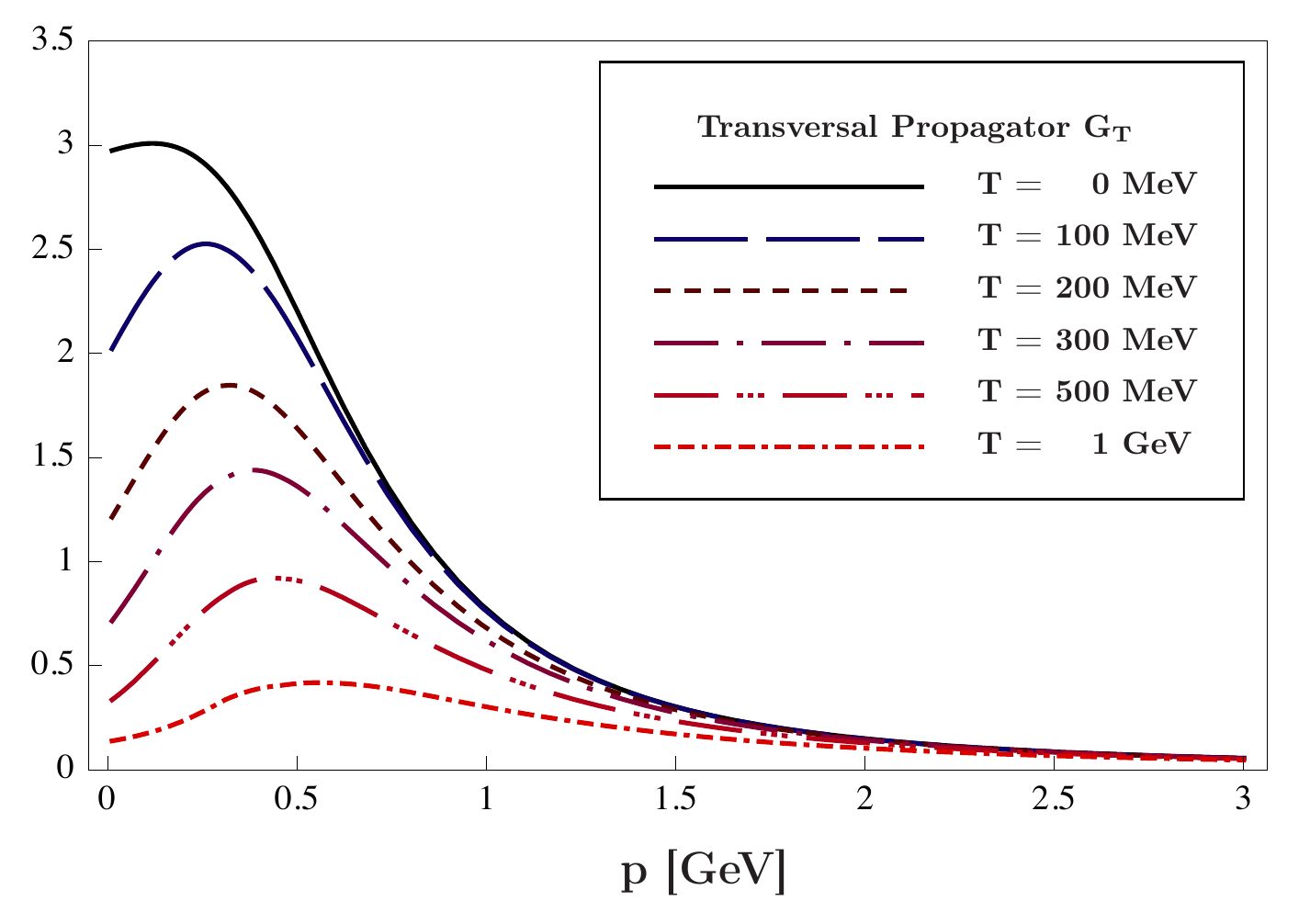}
\caption[]{Transversal gluon propagator $G_T$ at different
  temperatures as a function of spatial momentum.}
\label{fig:GT}
\end{figure}

\begin{figure}[t]
\includegraphics[width=.9\columnwidth]{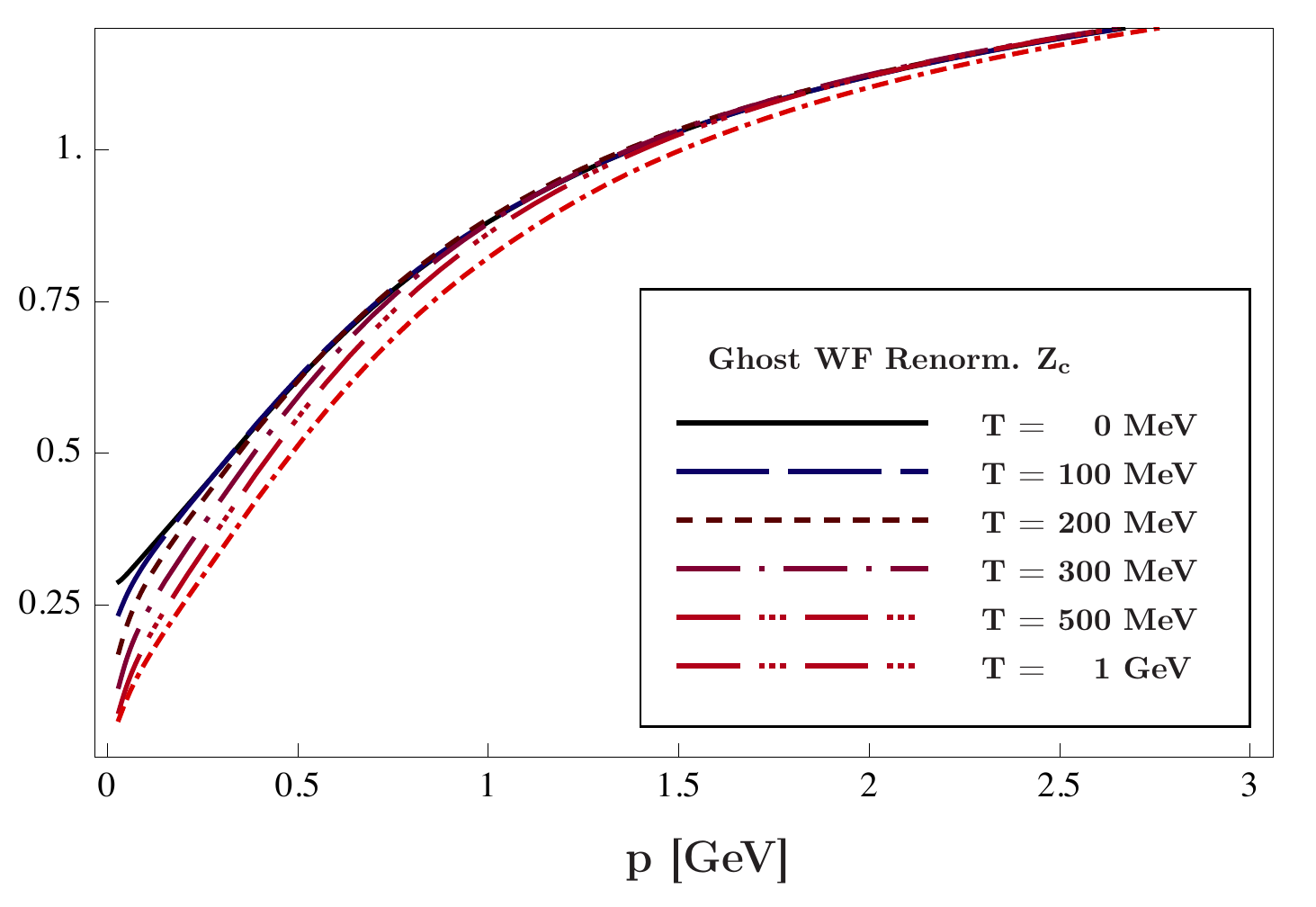}
\caption{Zero mode of the wave-function renormalisation of the ghost
  $Z_c$ at different temperatures as a function of spatial momentum.}
\label{fig:ZC}
\end{figure}

The transversal mode is not enhanced for small temperatures, in clear
distinction to the longitudinal mode. It is monotonously decreased with
temperature, see Fig.~\ref{fig:GT}.  Moreover, it develops a clear
peak at about 500 MeV. This can be linked to positivity violation,
which has to be present for the transversal mode: in the high
temperature limit it describes the remaining dynamical gluons of
three-dimensional Yang-Mills theory in the Landau gauge. The infrared
bending is more pronounced as that of respective lattice results, and
its strength is subject to the lack of quantitative precision at these
scales. In turn, for larger momenta the transversal propagator agrees
well with the respective lattice propagator. 
The ghost only shows a small temperature dependence, in
contradistinction to the gluonic propagators. This is fully compatible
with the lattice results, see 
\cite{Maas:2011ez,Aouane:2011fv,Maas:2011se,Cucchieri:2011di,%
Fischer:2010fx,Cucchieri:2007ta}. 
\begin{figure}[t]
\includegraphics[width=.9\columnwidth]{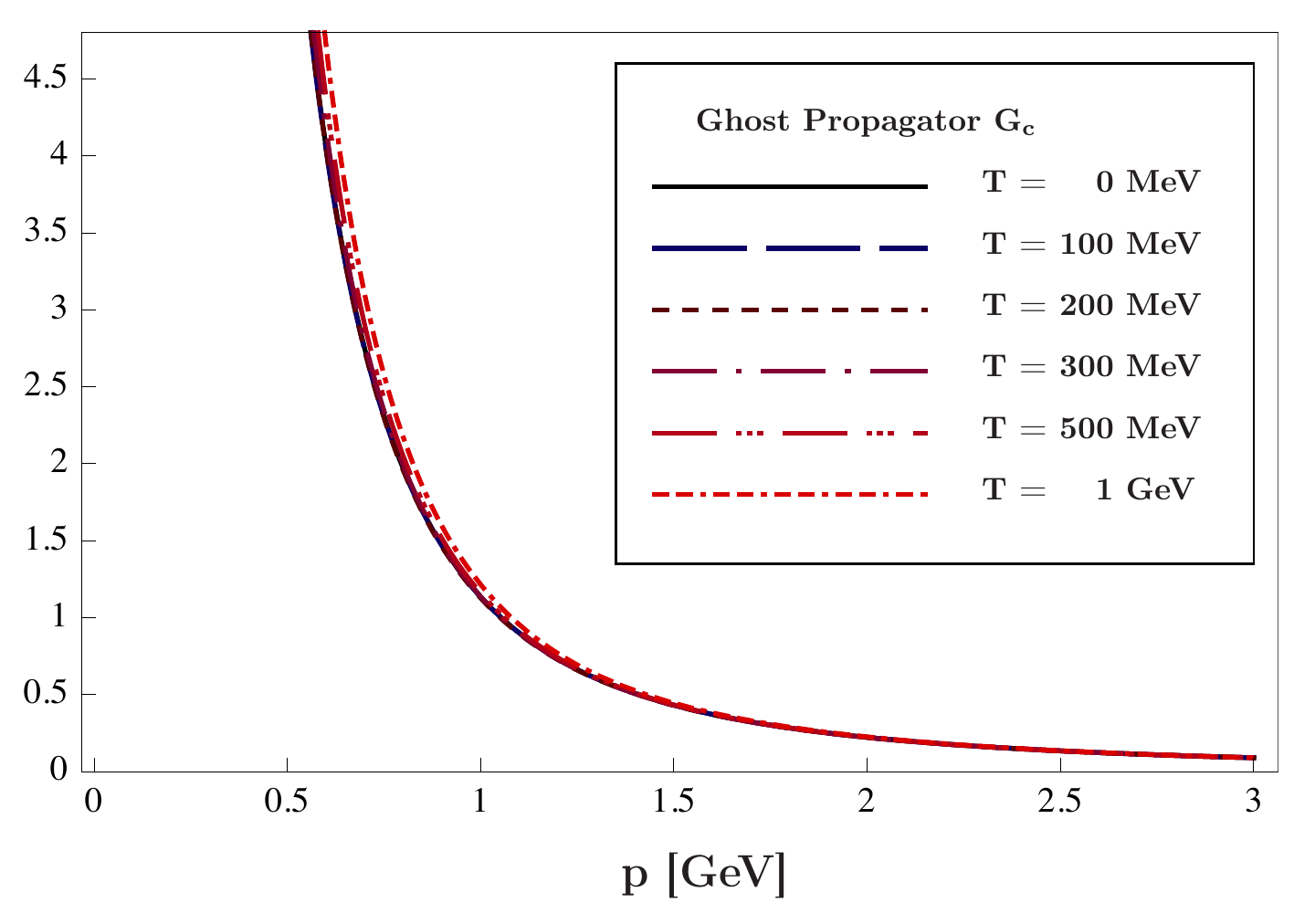}
\caption{Ghost propagator $G_c$ at different temperatures as a
  function of spatial momentum.}
\label{fig:GC}
\end{figure}
The temperature
dependence is hardly evident on the level of the propagator, see
Fig.~\ref{fig:GC}, but can be resolved on the level of the
wave-function renormalisation, see Fig.~\ref{fig:ZC}. The wave-function
renormalisation is slightly suppressed, which corresponds to a
successive enhancement of the ghost propagator at finite temperature.
The enhancement of the ghost propagator is potentially self-amplifying,
as it feeds back into the flow of the ghost two-point function, see
Fig.~\ref{fig:truncation}. This would
cause a pole in the ghost propagator at some temperature, if the
ghost-gluon vertex did not change. However, in this case the flow of
the latter is dominated by the diagrams with two ghost propagators, see
Fig.~\ref{fig:truncation}, and
the ghost-gluon vertex flows to zero. 

This non-trivial interplay of the flow of the ghost propagator with
the flow of the ghost-gluon vertex leads to a self-stabilising system
and prevents a further enhancement of the ghost.  This effect is
crucial for the stability of the solution of the Yang-Mills system at
finite temperature. Indeed, for a constant vertex no solution could be
obtained for intermediate temperatures $T\approx T_c$ and above. Thus
we conclude that any reliable truncation must comprise direct thermal
effects also in $n$-point functions with $n\leq3$.

\begin{figure}[t]
\includegraphics[width=.9\columnwidth]{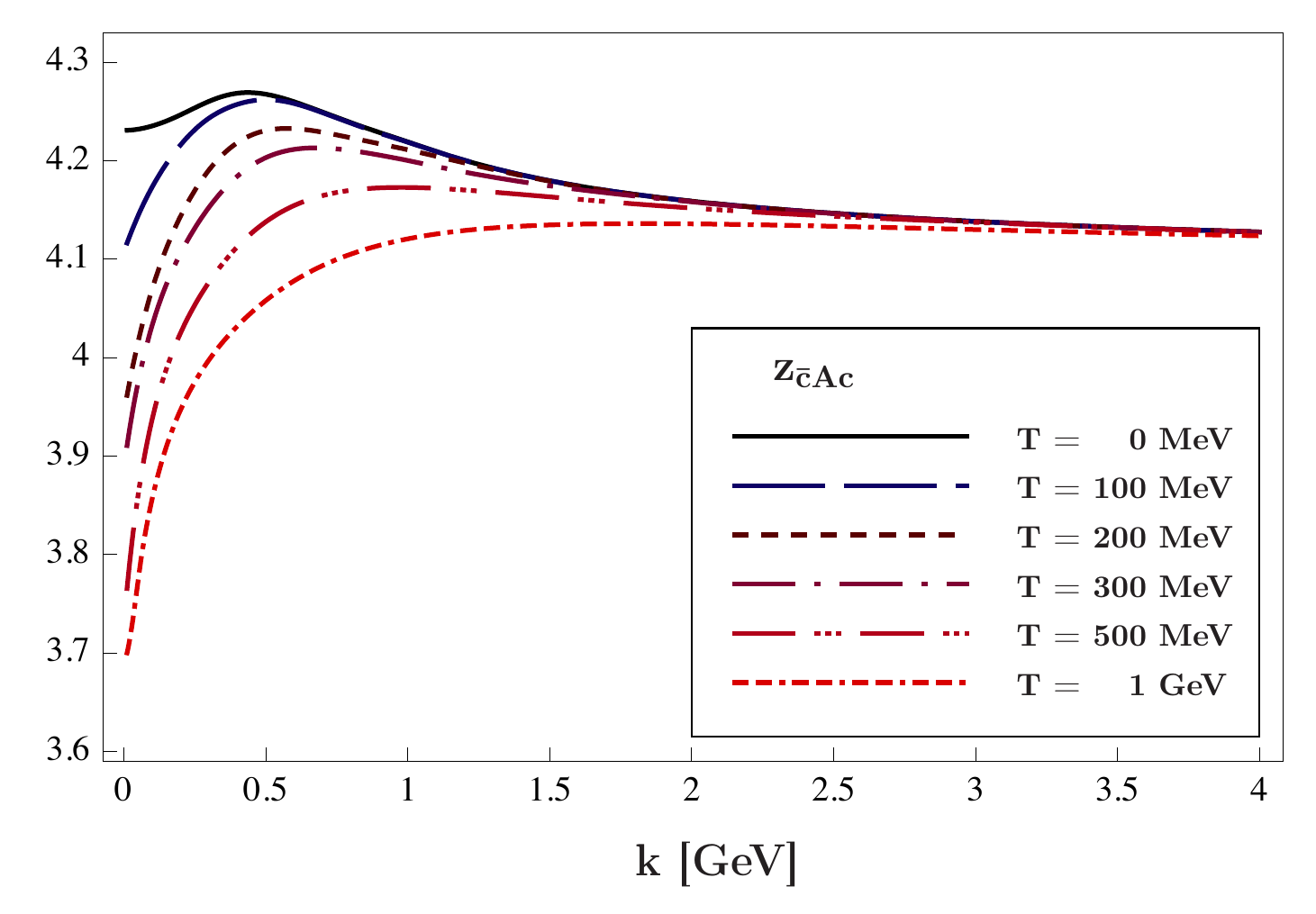}
\caption{Dressing function $z_{\bar{c}Ac}$ of the ghost-gluon vertex
  at different temperatures as a function of spatial momentum.}
\label{fig:ZcAc}
\end{figure}

The self-stabilising property of the Yang-Mills system explained above
is clearly seen in the temperature-dependence of the ghost-gluon
vertex.  The vertex is suppressed successively with the temperature,
see Fig.~\ref{fig:ZcAc}, which in turn ensures the relatively mild
change of the ghost propagator. Especially the sharp drop-off of the
vertex at small scales $k$ accounts for the smallness of the thermal
fluctuations to the ghost propagator.

\begin{figure}[t]
\includegraphics[width=.9\columnwidth]{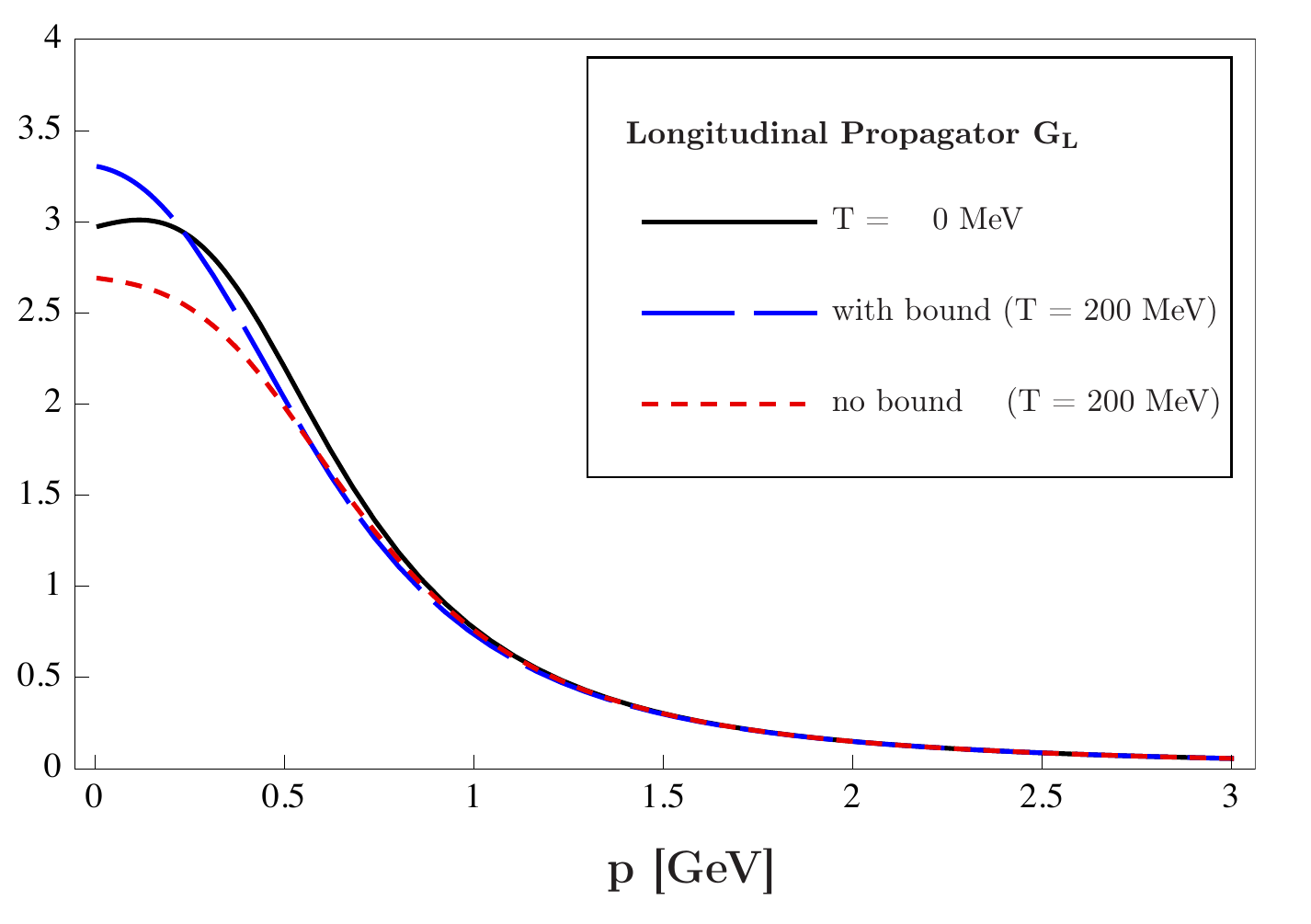}
\caption{Comparison of longitudinal propagator with minimal and
  maximal strength of the coupling of purely gluonic vertices.}
\label{fig:SO_GL}
\end{figure}

In Section~\ref{subsec:gluonicvertices}, we approximated the
temperature dependence of the gluonic vertices by an ansatz that
combined all necessary qualitative properties of the vertices. In
order to check the sensitivity of the propagators at low temperatures
to this choice, we compare the results with a strong suppression with a
computation where the minimal suppression of the gluonic vertices
defined in eq.~(\ref{eq:finaldressing}) is implemented. This gives an
approximate error band for the propagators with respect to the ansatz
of the gluonic vertices. We find that the main impact of the
truncation of the gluonic vertices is onto the longitudinal
propagator below $T_c$, given in Fig.~\ref{fig:SO_GL}. 
We have already mentioned that our truncation has to be improved in
order to quantitatively capture the deep infrared, and the temperature
dependence of the gluonic vertices is an important ingredient. 

\begin{figure}[t]
\includegraphics[width=.9\columnwidth]{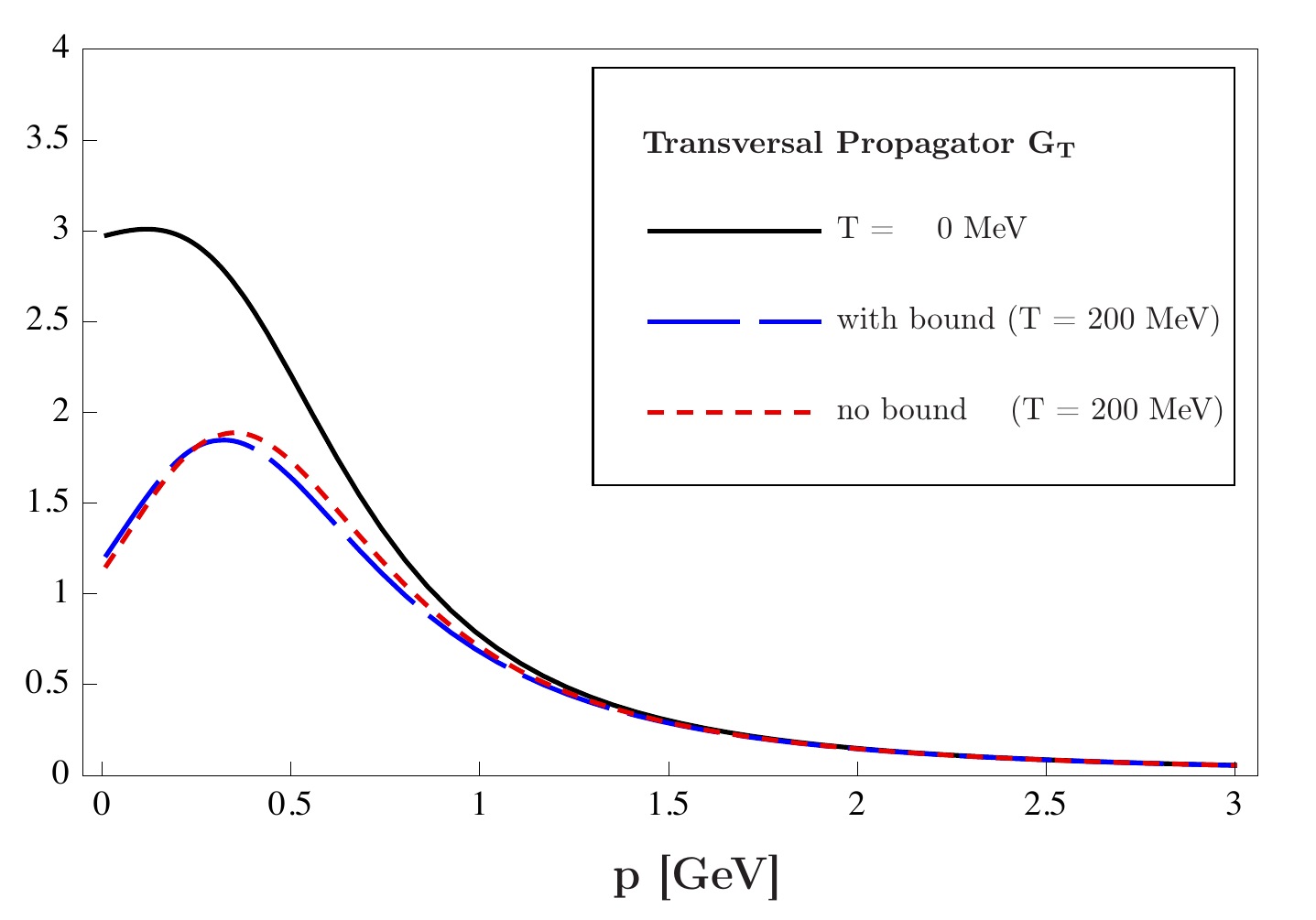}
\caption{Comparison of transversal propagator with minimal and maximal
  strength of the coupling of purely gluonic vertices.}
\label{fig:SO_GT}
\end{figure}

In contrast to the sensitivity of the deep infrared behaviour of the
longitudinal gluon, the transverse gluon propagator, see
Fig.~\ref{fig:SO_GT}, as well as the ghost propagator hardly feel the
modified infrared behaviour of the gluonic $n$-point functions.
This is in agreement with the well-known fact that the infrared sector
of Yang-Mills theory in Landau gauge has ghost dominance (for both
scaling and decoupling solutions) for any dimension $d=2,3,4$.  

\begin{figure}[t]
  \includegraphics[width=.9\columnwidth]{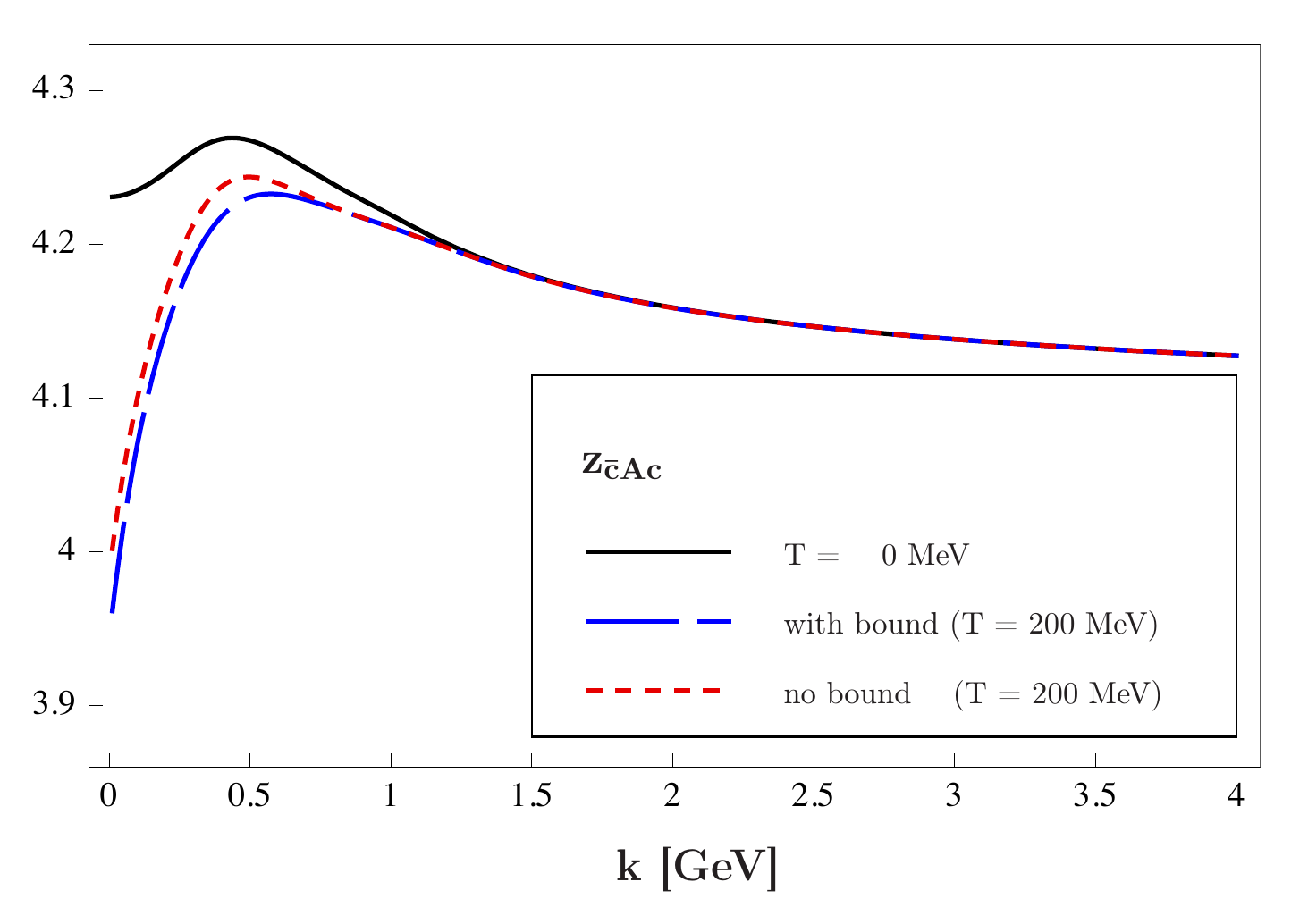}
  \caption{Comparison of ghost-gluon vertex with unbounded and bounded
    gluonic flow.}
\label{fig:SO_ZcAc}
\end{figure}

As the wave-function renormalisation of the ghost with the minimal and
maximal cut off of gluonic vertices can not be distinguished by eye,
we desist from illustrating it explicitly. In contrast to this, the
ghost-gluon vertex reflects the switching off of gluonic vertices, see
Fig.~\ref{fig:SO_ZcAc}, however, only in a region where the gluonic
contributions to the flow are subleading.

In summary, the scheme-dependence of the gluonic vertices affects the
details of the thermal decoupling of the longitudinal propagator, but
has no effects on the transversal propagator and the ghost. The
sensitivity of the thermal decoupling of the propagator to the thermal
decoupling of the vertices is easily understood. In turn, transversal
gluon and ghost are the relevant degrees of freedom for asymptotically
high temperatures where we are led to a three-dimensional confining
theory. This property can be seen nicely on the lattice, where only the
transversal gluon is sensitive to the removal of confining
configurations, see e.g.  \cite{Chernodub:2011pr}. We conclude that
our approximations do not affect the confining physics.

\begin{figure}[t]
\includegraphics[width=.9\columnwidth]{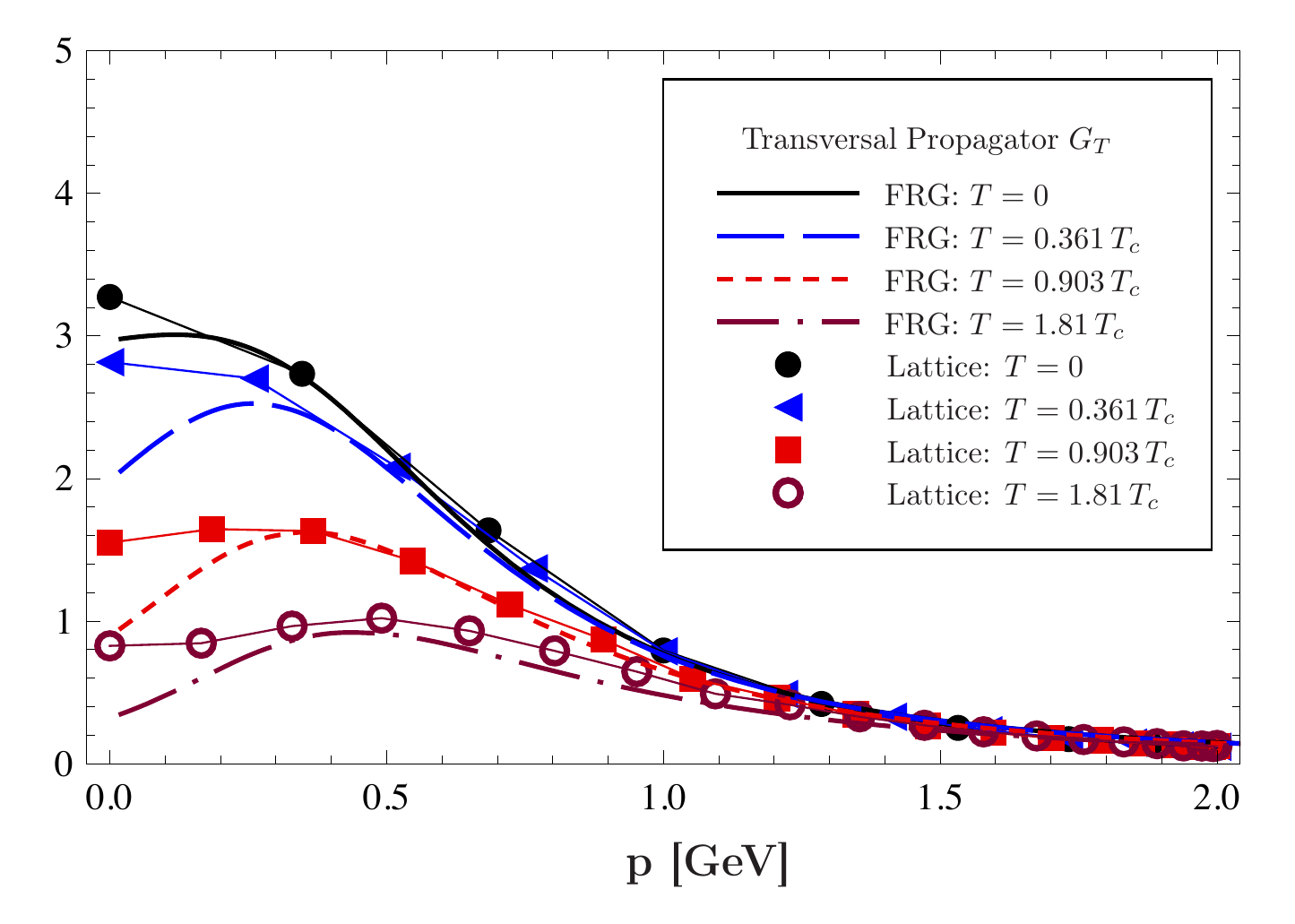}
\caption[]{Transversal gluon propagator in comparison with lattice
  results \cite{Maas:2011ez,Fischer:2010fx}. The lattice data has been
  rescaled such that the $T=0$ propagators match at intermediate
  momenta $p\gtrsim 1\tn{ GeV}$.}
\label{fig:FRG_latt_trans}
\end{figure}

In the following, we compare the propagators above with lattice
results \cite{Maas:2011ez,Fischer:2010fx, AxelPrivComm}. For this
purpose, we scale the lattice data such that the lattice propagators
at vanishing temperature match our normalisation at momenta $p\gtrsim
1\textnormal{ GeV}$. Take notice that we did not use the lattice
propagator as the initial condition, thus the deep infrared of the
data deviates from our propagator already at zero temperature, which
persists also in the propagators at finite temperature. Apart from
that, there is quantitative agreement with the lattice data with
respect to the (temperature dependent) momentum region, where the
thermal effects appear. In Fig.~\ref{fig:FRG_latt_trans} the
transversal propagators are compared. The critical temperature in the lattice data is $T_c=277\,\tn{MeV}$.
Clearly, we match the lattice propagator, except for the strong
bending of the FRG propagator in the infrared region. This difference
is a direct consequence of the mismatch of the ghost and gluon
propagators in the deep infrared at vanishing temperature, and relates
to the different decoupling solution chosen here. Apart from the deep
infrared, the momentum and temperature behaviour of the magnetic gluon
matches that of the lattice propagator. In contrast to this, the
electric gluon on the lattice shows a qualitatively different
behaviour for temperatures below and around the phase
transition. Although the longitudinal propagators agree for
$T=0.361T_c\approx 100\tn{MeV}$, it is exactly this region where the
uncertainty due to the truncation for the gluonic vertices is large,
as shown in Fig.~\ref{fig:SO_GL} for $T=200$ MeV.
Being aware of a potential truncation dependency in the deep infrared
of the longitudinal propagator at low temperatures, we note that in
the present truncation the electric gluon shows the onset of the
enhancement found on the lattice. Increasing the temperature this
feature disappears, and we see a qualitatively different effect for
temperatures below $T_c$. While the continuum result shows a strictly
monotonic decreasing propagator, the counterpart on the lattice is
enhanced in the confining regime, but reflects the phase transition in
form of a rapid decrease at $T_c$. Nevertheless, this deflection is
expected to be missed in the present truncation, as the Polyakov loop
potential $V(A_0)$ is pivotal for the critical behaviour around the
phase transition. In a full calculation the inverse propagator is
proportional to the second derivative of the Polyakov loop
$\Gamma_{A,L}^{(2)}\sim V''(A_0)$, see
\cite{Pawlowski:2010ht,Braun:2010cy,Braun:2009gm,Braun:2007bx}.  This
dependence introduces an additional screening of the
vertex-approximations as well as vertex corrections, but in the
computations presented here it was dropped. This upgrade should also
give us access to the question of the signatures of criticality in the
chromoelectric propagator discussed in \cite{Maas:2011se}.
Interestingly, such a term is absent in the magnetic modes, which do
not depend strongly on the vertex-approximation. It is suggestive that
the inclusion of $A_0$ will further stabilise our
computation. However, we defer this interesting computation of the
propagator in the presence of a non-trivial background to future work.

\section{Summary and outlook}\label{sec:summary}

In the present work we have put forward a functional approach for the
quantitative computation of full, momentum dependent correlation
functions at finite temperature in Yang-Mills theory. This is done
within a FRG-setting which only incorporates thermal fluctuations. 

We have computed temperature dependent Yang-Mills propagators and
vertices in the Landau gauge for temperatures $T\lesssim 3 T_c$. The
chromoelectric propagator shows the expected Debye-screening for
$T>T_c$ in quantitative agreement with the lattice results. For small
temperatures it shows qualitatively the enhancement also seen on the
lattice \cite{Maas:2011se,Cucchieri:2007ta,Fischer:2010fx%
  ,Bornyakov:2011jm,Aouane:2011fv,Maas:2011ez,Cucchieri:2011di}.
However, the significance of the lattice results so far as well as
quantitative details are not settled yet. The chromomagnetic
propagator shows the expected thermal scaling and tends towards the
three-dimensional gluon propagator in quantitative agreement with the
lattice. The ghost propagator only shows a mild enhancement with
temperature in agreement with the lattice. In contradistinction we see
a strong thermal infrared suppression of the ghost-gluon vertex which
increases with temperature. The results of this paper are also
summarised in \cite{proceedings}. 

At present we improve on the approximations made here and compute
thermodynamic observables such as the pressure and the scale anomaly.
The inclusion of the $A_0$-corrections discussed in
Section~\ref{sec:props+vertices} should give access to the properties 
of the chromoelectric propagator at criticality, see \cite{Maas:2011ez}. 
Furthermore, we utilise the
propagators obtained here in a combined approach with FRG and DSE as
well as lattice results in an extension of \cite{Fischer:2009wc}. This
work aims at the QCD phase structure for heavy quarks, for lattice
results see \cite{Philipsen:2011zx,Fromm:2011qi}. Moreover, the glue
and ghost propagators are a key input for quantitative computations in
full QCD with 2 and 2+1 flavours at finite temperature and density in
an extension of \cite{Braun:2009gm}. \step

\noindent {\it Acknowledgements} \\[0ex]
We thank R.~Alkofer, J.-P.~Blaizot, J.~Braun, S.~Diehl, C.~F.~Fischer,
K.~Fukushima, L.~M.~Haas, M.~Ilgenfritz, A.~Maas, U.~Reinosa,
B.-J.~Schaefer, A.~Sternbeck, L.~von Smekal and C.~Wetterich for
discussions, M.~Ilgenfritz, A.~Maas, A.~Sternbeck for providing
lattice results. This work is supported by the Helmholtz Alliance
HA216/EMMI.  LF acknowledges financial support by the Helmholtz
International Center for FAIR within the LOEWE program of the State of
Hesse and the Helmholtz Young Investigator Grant VH-NG-322.

\noindent 

\bibliographystyle{bibstyle}
\bibliography{bib}


\end{document}